\documentclass[manuscript]{rmaa}
\usepackage{graphicx}
\usepackage{paralist}
\usepackage{psfrag,color}
\usepackage{longtable}

\usepackage{array}
\newcommand{\PreserveBackslash}[1]{\let\temp=\\#1\let\\=\temp}
\newcolumntype{C}[1]{>{\PreserveBackslash\centering}p{#1}}
\newcolumntype{R}[1]{>{\PreserveBackslash\raggedleft}p{#1}}
\newcolumntype{L}[1]{>{\PreserveBackslash\raggedright}p{#1}}

\title{Photomeric analysis of eclipsing binaries: VY UMi, RU UMi and GSC 04364-00648}

\author{
  V. Kudak,\altaffilmark{1} 
  \v{S}.Parimucha,\altaffilmark{2}
  V. Perig,\altaffilmark{1} 
  and P. Gajdo\v{s}\altaffilmark{2} }

\altaffiltext{1}{Laboratory of space researches, Uzhhorod National University, Uzhhorod, Ukraine.}
\altaffiltext{2} {Institute of Physics, Faculty of Science, P.J. \v{S}af\'arik University, Ko\v{s}ice, Slovakia.}

\shortauthor{Kudak et al.}
\shorttitle{Analysis of 3 neglected binaries}

\fulladdresses{
\item V.Kudak and V.Perig: Laboratory of space researches, Uzhhorod National University, Uzhhorod, Daleka Str., 2A,  88000, Ukraine (lab-space@uzhnu.edu.ua)

\item M.Fedurco, P. Gajdo\v{s} and \v{S}.Parimucha: Institute of Physics, Faculty of Science, P.J. \v{S}af\'arik University, Ko\v{s}ice, Park Angelinum 9, 04001, Slovakia (stefan.parimucha@upjs.sk)
 }

\listofauthors{V.Kudak, \v{S}.Parimucha, M.Fedurco, V.Perig, \& P. Gajdo\v{s}}
\indexauthor{Kudak, V.}
\indexauthor{Parimucha, \v{S}.}
\indexauthor{Fedurco, M.}
\indexauthor{Perig, V.}
\indexauthor{Gajdo\v{s}, P.}

\abstract{
We present the photometric analysis of \textit{BVR} and \textit{TESS} light curves of three eclipsing binaries (RU~UMi and purely studied VY~UMi, GSC 04364-00648), together with their period changes considering archival data and new minima times from our and \textit{TESS} observations. For the first time we detected wave-like variations with low-amplitude
in $O-C$ residua of RU UMi, which can be interpreted as a consequence of the
light-time effect caused by the 3rd invisible component with period 7370 days.
Period increase with rate 2.56(9)$\times10^{-7}$~d/yr$^{-1}$ detected in the VY UMi
system corresponds to mass transfer from the secondary to the primary component.
For GSC 04364-00648 binary system we find some quadratic changes on the $O-C$ diagram, which corresponds
to a period decrease with a high rate of $-2.26(5)\times10^{-5}$~d/yr$^{-1}$. We cannot 
assumptions about their nature, mainly due to short time of observation and uneven coverage of $O-C$ diagram.
We also determined the absolute parameter of their components using the photometric solution and \textit{GAIA} distances.
}

\resumen{}

\addkeyword{Binaries: close}
\addkeyword{Binaries: eclipsing}
\addkeyword{Stars: Mass loss}
\addkeyword{Stars: individual (RU UMi, VY UMi, GSC 04364-00648)}

\begin{document}
\maketitle

\section{Introduction}
\label{sec:intro}

Eclipsing binary stars are systems, where the components are mutually obscured for the
observer during their motion around a common centre of mass. It is a very important group of variable stars with specific and well-recognized light-curves, where their shapes depend on the physical properties of the components and geometrical configuration  \citep[]{2001icbs.book.....H, Prsa2019, 2021A&A...652A.156C}. 

Analysis of light-curves of eclipsing binaries can reveal, among other aspects, the relative dimensions of stars, their effective temperatures, orbital inclination, the eccentricity of the orbit, and potential spots on their surfaces. Together with radial velocities obtained from spectroscopic observations, we can determine the masses of the components, their radii and luminosities, and the dimension of their orbit. These parameters can be also estimated if we know the distance and the amount of interstellar extinction to star.

The shapes of components in binary stars are described by Roche geometry \citep[e.g][]{Prsa2019}. According to this, three configurations of binaries are possible, detached (both components are in their Roche lobes), semidetached (one component fills its Roche lobe), and contact, where both components overfill their Roche lobes. All this is reflected in the light-curves and also has other observational consequences, like a period change due to mass transfer, angular momentum loss \citep[e.g.][]{Yang2009} and/or magnetic braking \citep{1992ApJ...385..621A}.

We now know more than 500\,000 eclipsing binaries \citep[]{2006SASS...25...47W}, and in the era of large photometric surveys \citep[e.g.][]{2019ApJ...873..111I} one can expect the discovery of several million new eclipsing binaries. But only a very small fraction of them (less than 1\%) have calculated parameters. It is a big challenge for data analysis in the near future. 

In this paper, we want to make a small contribution to the knowledge of eclipsing binary stars by photometry study and period analysis of three eclipsing binaries, 
two of them were not studied in detail up to now in literature, while for 3rd we indicate the possible presence of 3rd body according to the $O-C$ diagram. 

\textbf{RU UMi} (TYC 4402-504-1) was for the first time mentioned as eclipsing binary of Algol type by \citet{1968BamVe...7...72S}. They analysed sky-patrol plates taken from 1931 through 1959. \citet{1971PASP...83..286W} presented the first photometric solution and concluded, that both stars are oversize for their masses and the object may be a contact system of W UMa type. Other photometric solutions by \citet{1973AJ.....78..107N} and \citet{1977AcA....27..187D} suggested that the system was close to contact, while \citet{1985AcA....35..327K} 
modeled previous data and concluded that the system was quite close to a semidetached configuration. It was supported by \citet{1988PASJ...40...79O},  \citet{1993MNRAS.260..478B} and  \citet{2006PASJ...58..361Z}. The radial velocities for the primary component were published by \citet{1988PASJ...40...79O} and for both components by \citet{1996Obs...116..288M}. They found that the mass ratios in the range $0.32<q<0.40$ provide a good solution to the light curves.

\citet{2006PASJ...58..361Z} published $O-C$ period analysis of up to date minima times observations and suggested a continuous period decrease at a rate $dP/dt =-1.72 \times 10^{-8}$ d\,yr$^{-1}$ caused by a transfer of the matter from the secondary to the primary component.  \citet{2008PASP..120..720L} explained the secular period decrease by angular momentum loss (AML) due to magnetic braking alone or more convincingly by a combination of AML and mass transfer from the less massive secondary to the more massive primary. The distance to the system is 283.0$\pm$1.2 pc \citep{2022arXiv220605989B}. 
RU~UMi was observed in 5 sectors during \textit{TESS} mission \citep{2014JAVSO..42..234R}.



\textbf{VY UMi} (GSC 04568-00313) was discovered as variable star by \citet{1958Strohmeier}. The first ephemeris for this eclipsing binary was published by \citet{2004IBVS.5557....1O}. The distance to the system is 164.5$\pm$0.3 pc
\citep{2022arXiv220605989B}.
VY UMi has no published photometric solution of the light-curve and neither period analysis up to now. Meanwhile, the object was observed in 13 sectors during the \textit{TESS} mission, so it is an interesting object for our research.
 
\textbf{GSC 04364-00648} (TYC 4364-648-1) was mentioned as variable in Wide-field Infrared Survey Explorer (WISE) Catalog of Periodic Variable Stars by \citet{2018ApJS..237...28C} with period $0.8628506$ d. The distance to the system was determined to be 512.5$\pm$4.8 pc 
\citep{2022arXiv220605989B}.
The system has observations from 3 \textit{TESS} mission sectors. No other analysis of this eclipsing binary was published.


\section{Observations and data reduction}
All new CCD observations of eclipsing binary systems presented in this work were carried out at the Derenivka Observatory of Uzhhorod National University, Ukraine (Lat: 48.563 N; Long: 22.453 E, MPC code K99) and Kolonica Astronomical Observatory (KAO) of the P.J. \v{S}af\'arik University, Ko\v{s}ice, Slovakia (Lat: 48.950 N; Lon: 22.266 E). Measurements were collected from March 2021 to October 2021.

In Derenivka Observatory, we have used a 400 mm Newton-type telescope with a focal ratio of f/4.4 
equipped with FLI PL9000 CCD camera (array 3056$\times$3056, pixel size 12$\mu m$) with $BVR$ Bessel photometric filters. The field of view of such configuration of the system is $1.21\arcdeg \times1.21\arcdeg$.
Observations at KAO were made by PlaneWave CDK20 telescope with a diameter of main mirror 508 mm and a focal ratio of f/6.8 at the Cassegrain focus. The telescope is equipped with a G4-16000 CCD camera (array 4096$\times$4096, pixel size 9$\mu m$) with $UBVRI$ Bessel photometric filters. The field of view of the system is $37\arcmin \times 37\arcmin$. 
The detailed journal of our CCD observation is given in Table~\ref{tab:obs-log}.

\begin{table}[t]\centering
  \setlength{\tabnotewidth}{0.85\columnwidth}
  \tablecols{5}
  \setlength{\tabcolsep}{1.\tabcolsep}
  \caption{The journal of our CCD observations.} \label{tab:obs-log}
\small
\begin{tabular}{clccc}
\hline \hline
\textbf{System}     &\textbf{Date} & \textbf{Time(UT)}  &\textbf{Phase}\tabnotemark{a}  & \textbf{Filters} \\
\hline
RU UMi  & Mar 03 21 & 17:24 - 23:05  & 0.646 - 0.097 & $BVR$   \\
        & Sep 06 21 & 20:43 - 01:00  & 0.151 - 0.490 & $BVR$   \\
        & Sep 09 21 & 17:53 - 00:48  & 0.640 - 0.190 & $BVR$   \\
        & Sep 12 21 & 17:53 - 21:49  & 0.356 - 0.669 & $BVR$   \\
\hline
VY UMi     & Mar 24 21 & 18:36 - 03:35  & 0.729 - 0.878 & $BVR $\\ 
           & Oct 28 21 & 16:47 - 01:12  & 0.430 - 0.509 & $BVR$ \\ 
           & Oct 29 21\tabnotemark{b} & 18:50 - 03:03  & 0.767 - 0.819 & $BVR$ \\ 
\hline
GSC         & Apr 04 21 & 18:28 - 03:02  & 0.469 - 0.883 & $BVR$ \\ 
04364-00648 & Apr 08 21 & 18:27 - 23:34  & 0.104 - 0.351 & $BVR$ \\ 
            & Apr 10 21 & 18:28 - 02:47  & 0.422 - 0.824 & $BVR$ \\ 
            & May 08 21 & 19:11 - 01:52  & 0.908 - 0.230 & $BVR$ \\ 
            & Jun 08 21 & 20:00 - 00:51  & 0.875 - 0.109 & $BVR$ \\ 
\hline\hline
\tabnotetext{a}{\small Phase is calculated according to ephemeris determined in Section~\ref{section:period}.}
\tabnotetext{b}{\small Observation made at KAO} 
\tabnotetext{c}{\small Photometrical data are available from the first author upon
request.}
\end{tabular}
\end{table}

\begin{table}[!t]\centering
  \setlength{\tabnotewidth}{0.85\columnwidth}
  \tablecols{5}
  \setlength{\tabcolsep}{1.\tabcolsep}
  \caption{Stars used for a determination of artificial comparison stars.} \label{tab:comp_stars}
\small

\setlength{\tabcolsep}{1pt}
\small
\begin{tabular}{L{1.8cm}|C{2.7cm}|C{1.8cm}C{2cm}|C{1cm}C{1cm}C{1cm}}
\hline \hline
\textbf{System}     &   \textbf{Comparison stars}& \multicolumn{2}{c|}{\textbf{Coordinates}}&\textbf{\textit{B}}&\textbf{\textit{V}}&\textbf{\textit{R}}\\
\textbf{} &   \textbf{NOMAD}          &  $\alpha(2000)$ & $\delta(2000)$ &  &  & \\
\hline\hline
RU UMi      &1596-0121876&13:38:09.84&+69:41:12.9& 10.8230& 10.066& 9.580 \\
            &1596-0122007&13:40:43.97&+69:36:34.7&  9.769&  9.349& 9.070 \\
            &1599-0114087&13:42:10.79&+69:59:01.9& 10.410&  9.923& 9.590 \\
\hline
VY UMi &    1666-0084341&17:16:16.99&+76:37:00.25&11.725& 10.338& 9.470 \\
        &   1668-0086624&17:14:20.44&+76:52:34.29&11.313& 10.602& 10.130 \\
        &   1667-0084244&17:15:37.39&+76:42:45.44&12.036& 11.589& 11.290 \\
\hline
GSC         &1611-0075943&07:19:59.44&+71:08:30.09&12.288& 11.841& 11.540 \\
04364-0064  &1610-0077383&07:19:58.36&+71:05:32.56&11.157& 10.646& 10.310 \\
\hline \hline
\end{tabular}
\end{table}

The CCD images were calibrated (bias and dark subtraction, flat-field correction) utilizing the software package CoLiTecVS \citep{Savanevych2017, Parimucha2019}. This package was also used for aperture photometry, calculation of differential magnitudes according to artificial comparison star as well as calibration to the standard photometric system. The comparison stars used for the determination of artificial ones were selected manually according to the similarity of the studied binaries (brightness, distance in the sky). This approach significantly improves the quality of photometric measurements. 
Due to not a stable night-to-night observing conditions, the average precision of our measurements varied $\sim$0.01-0.05 mag in $V$ and $R$ filters and $\sim$0.03-0.08 mag in $B$ filter.
The comparison stars used in our study are listed in Table~\ref{tab:comp_stars}, together with their magnitudes from the NOMAD catalogue \citep{Zacharias2004, Zacharias2005}.

\begin{figure}[t]
	\centering
	\includegraphics[width=0.45\columnwidth, angle=0]{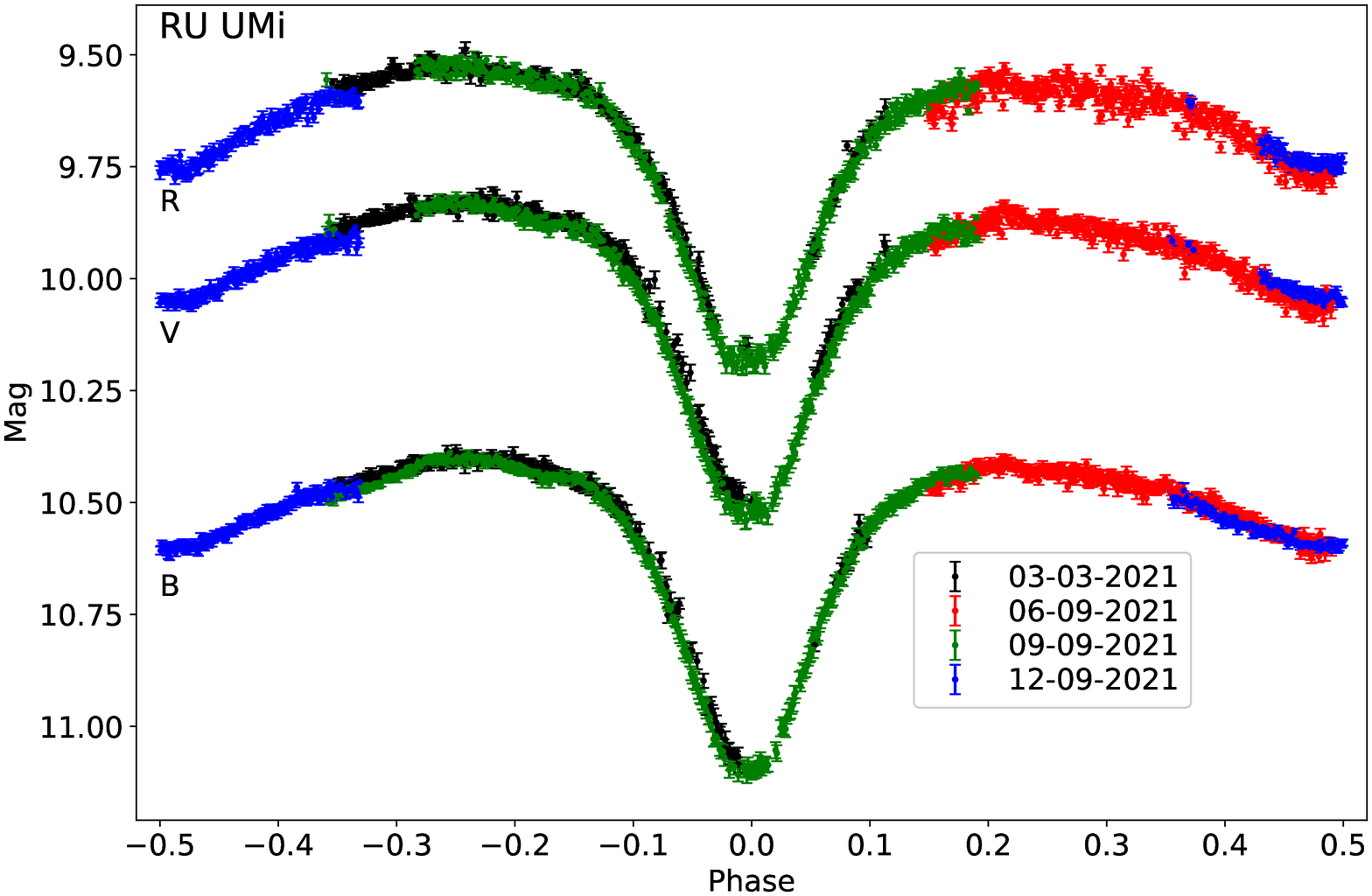}
	\includegraphics[width=0.45\columnwidth, angle=0]{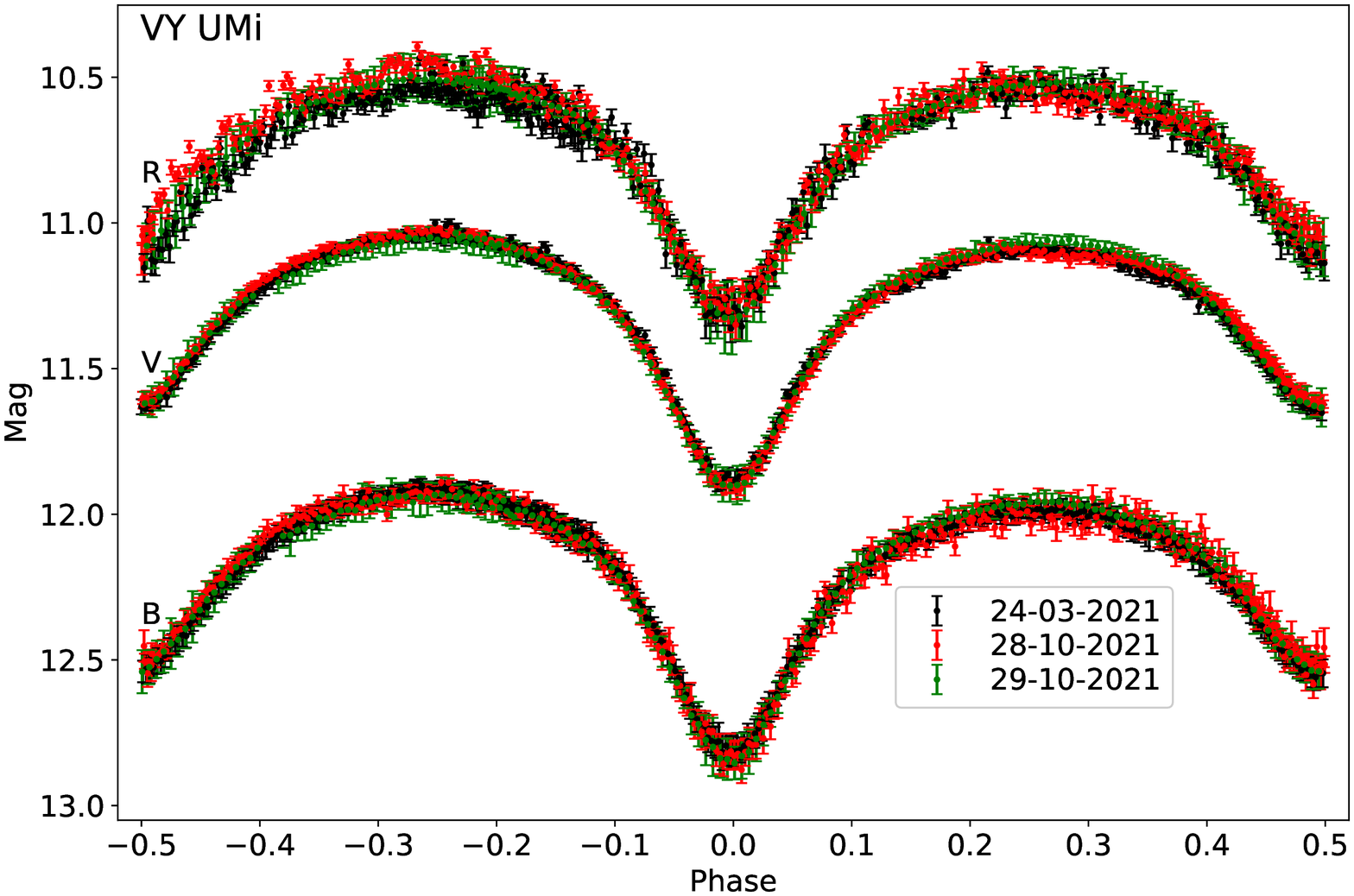}
	\includegraphics[width=0.45\columnwidth, angle=0]{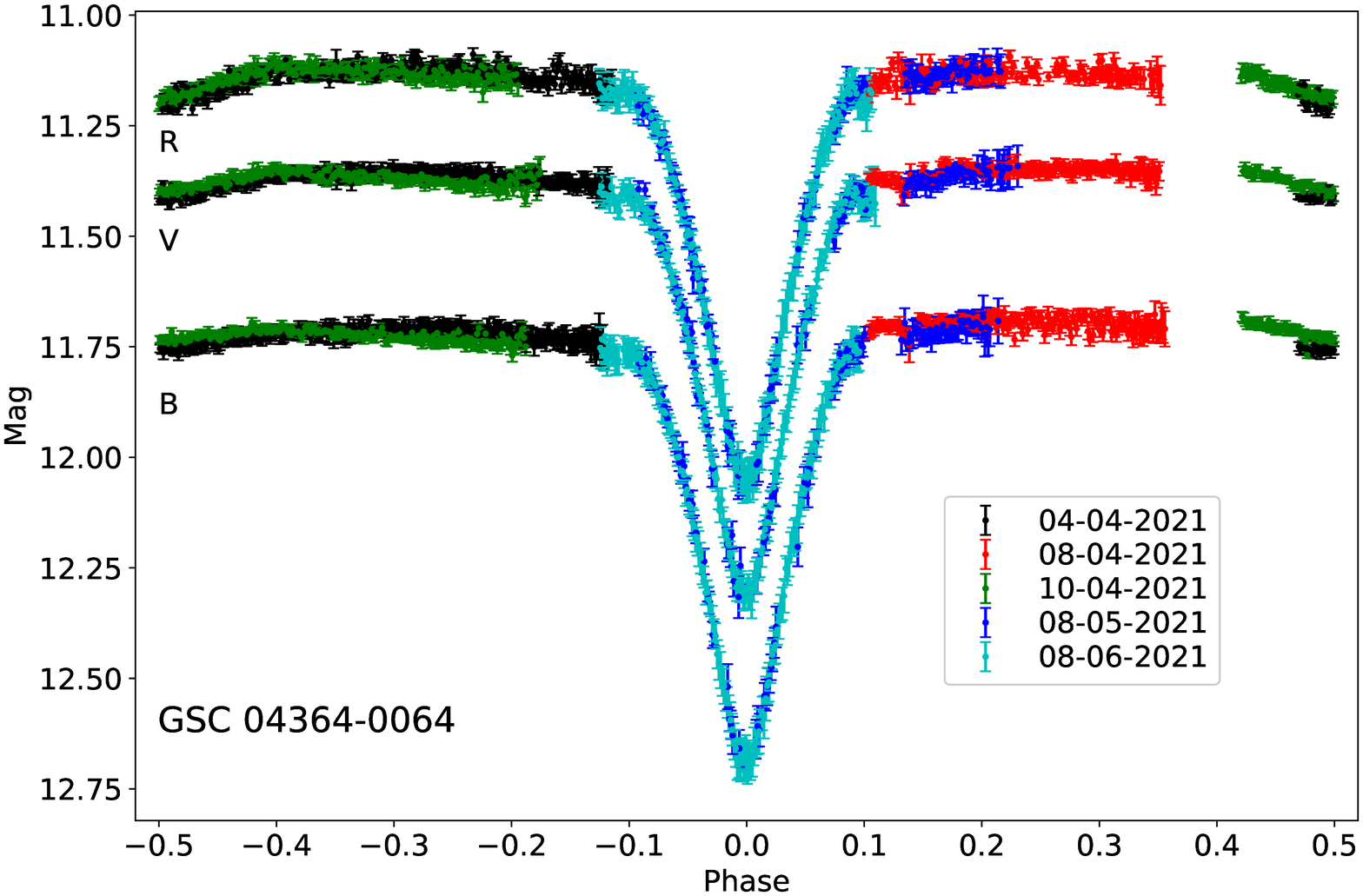}
	\caption{
	Phased light curves of RU UMi, VY UMi and GSC 04364-00648 in $BVR$ passbands by dates of observations. The phases are calculated according to ephemerides determined in Section \ref{section:period}}.
	\label{fig:phot_all}
\end{figure}

The resulting light-curves of all eclipsing binaries are depicted in Fig.~\ref{fig:phot_all}. The light-curves were phased according to ephemerides determined from $O-C$ variations analyzed in the next chapter. Magnitudes on the Fig.~\ref{fig:phot_all} have tiny systematic shifts according to APASS magnitudes that are compared to the level of observation errors.

\section{Analysis of period changes}
\label{section:period}

\begin{table}[t]\centering
  \setlength{\tabnotewidth}{0.85\columnwidth}
  \tablecols{4}
  \setlength{\tabcolsep}{1.\tabcolsep}
  \caption{Parameters of period changes detected in the studied eclipsing systems.} 
  \label{tab:oc_param}
\small
\begin{tabular}{cccc}
\hline \hline
        &\textbf{RU UMi} & \textbf{VY UMi}  &\textbf{GSC 04364-00648} \\
\hline
$P$[d]  &      --        & 0.3254004(4)     &  0.862839(1)  \\
$T_0$   &      --        & 2452500.0469(38) &  2459309.7273(2) \\
$Q$[d]  &      --        & 1.14(3)$\times10^{-10}$   &  -2.67(2)$\times10^{-8}$  \\
\hline
$P_3$[d]  &  7370(42)    &  --  &  -- \\
$T_{0_3}$   &  2458984(55) &  --  & --  \\
$e_3$     &  0.199(1)    &  --  &  -- \\
$\omega_3$[\arcdeg] &  116(3)    & --   & --  \\

$a_{12}\sin i_3$ [AU]  &  0.247(2)    &   -- &   -- \\
$f(m_3)$ [M$\odot$]  & 3.69(8)$\times10^{-5}$    &  --  &  --  \\

\hline\hline
\tabnotetext{}{\small $P$, $T_0$, $Q$ are period, reference minimum time and quadratic term, $P_3$, $T_{0_3}$, $e_3$, $\omega_3$, $a_{12}\sin i_3$ and $f(m_{3})$ describe parameters of the 3$^{\rm rd}$ body orbit, period, time of periastron passage, eccentricity, argument of periastron, projection of semi-major axis and function of the mass.}
\end{tabular}
\end{table}

In our analysis of period changes of all systems we have considered all available published minima
times as can be found in the $O-C$ gateway\footnote{http://var2.astro.cz/ocgate/} as well as minima times from our (weighted averages from $BVR$ light curves) and \textit{TESS} observations. 

Our new minima times were calculated following the phenomenological method described in \citet{2015A&A...584A...8M}. This method gives a realistic and statistically significant error in determining minima times. Newly calculated minima times from our and \textit{TESS} observations are listed in Tables \ref{tab:tmin_RU_UMi}-\ref{tab:tmin_our}.

\begin{figure}[t]
	\centering
\includegraphics[width=0.45\columnwidth, angle=0]{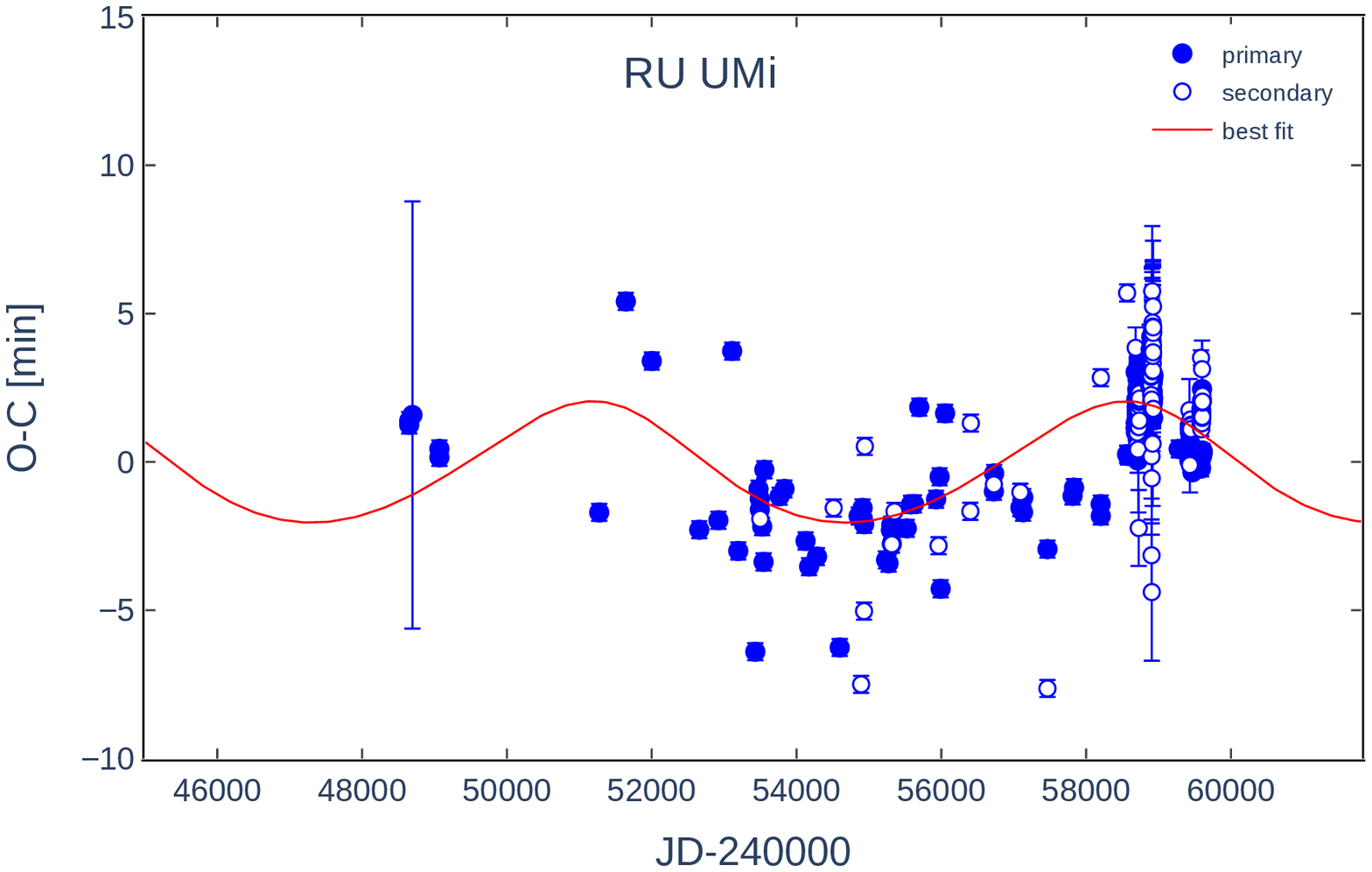}
\includegraphics[width=0.45\columnwidth, angle=0]{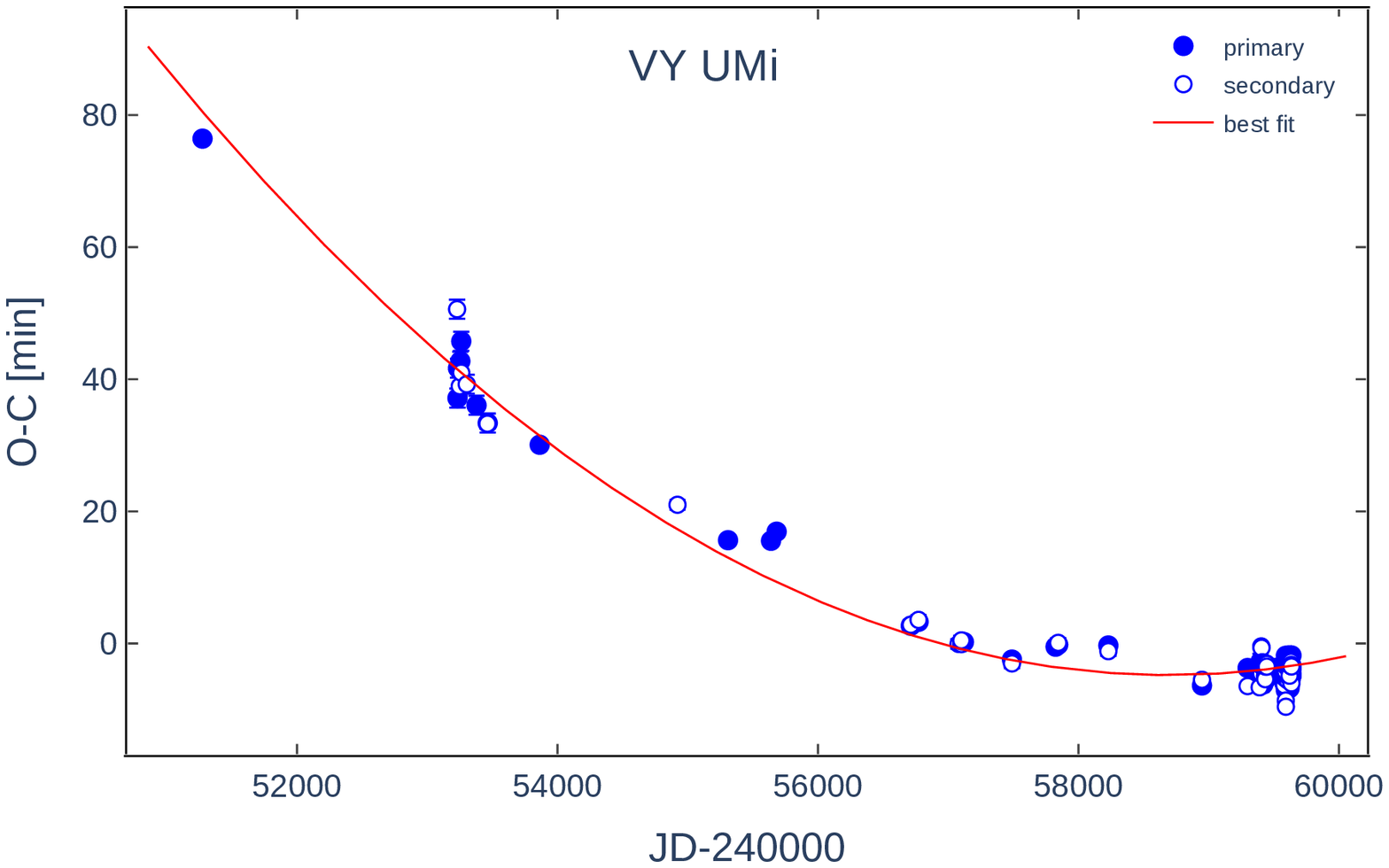}
\includegraphics[width=0.45\columnwidth, angle=0]{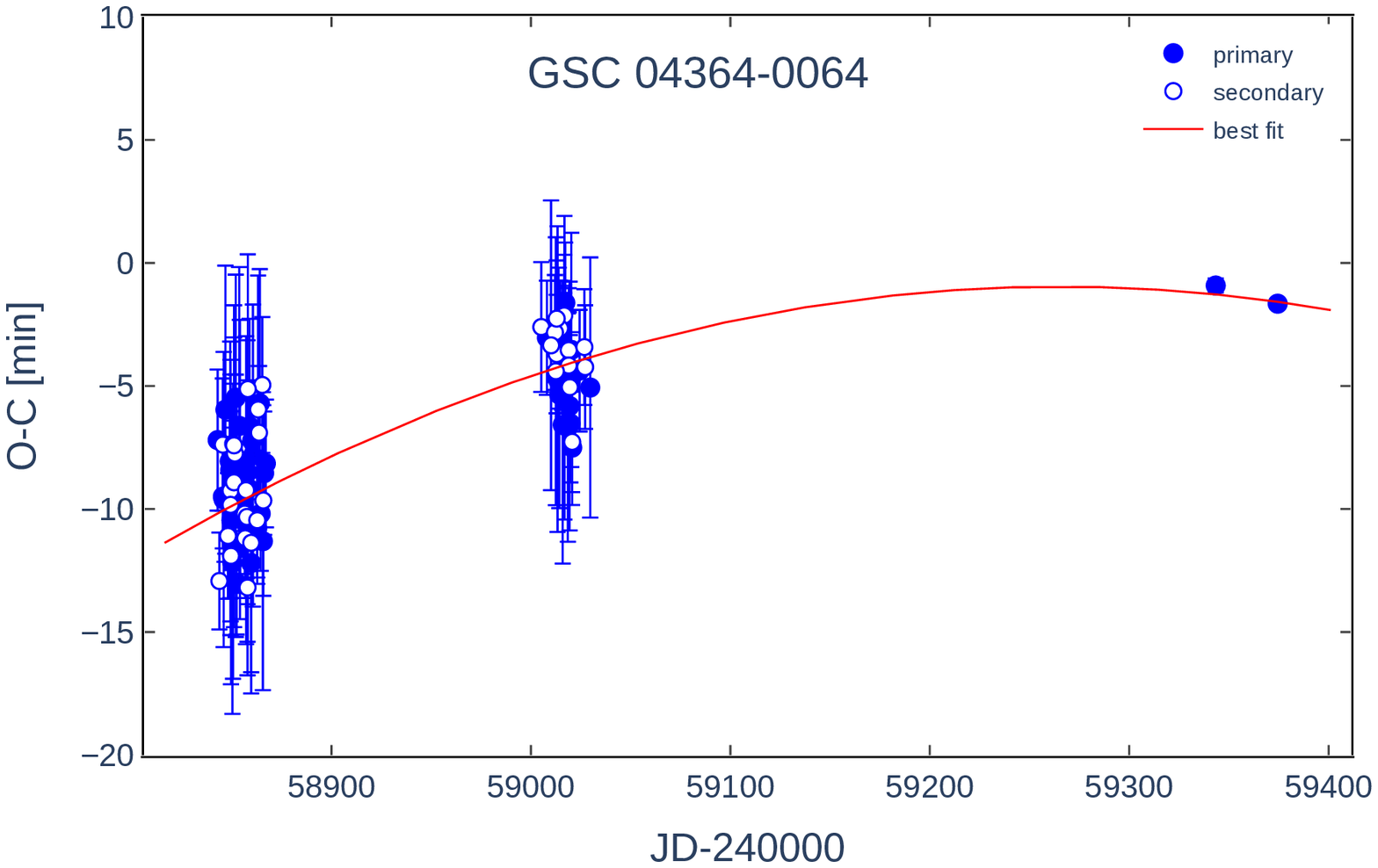}
	\caption{$O-C$ diagrams of studied systems according to linear ephemeris in Section~\ref{section:period}.}
	\label{fig:oc_all}
\end{figure}

Although the first minima times of RU UMi were obtained at the beginning of the previous century and cover almost the whole observed range, many of them are useless for detailed analysis. Photographic and visual estimates have a large scatter and one can expect their large internal errors. We decided to use only archived photoelectric and CCD minima times obtained since 1990 and minima times determined from our CCD and \textit{TESS} light curves. Their precision is in the range of  $10^{-4}$ days. A weighted least-squares solution using all selected minima (weights were calculated as $1/\sigma^2$, where $\sigma$ is a published or determined error of the minimum) leads to the following linear ephemeris of the RU UMi system (errors of parameters are given in parenthesis):
\begin{equation}
  \mathrm {Min~I} = \mathrm {HJD~} 2452500.0931(4) + 0^{d}.52492591(9) \times E.  
  \label{eq:lin_ruumi}
\end{equation}
This ephemeris was used to create the $O-C$ diagram displayed in Fig.~\ref{fig:oc_all}. The inspection of $O-C$ residua uncover their low-amplitude, wave-like variations.
It can be explained by the light-time effect caused by another invisible body orbiting a common center of the mass. To find the parameters of this orbit, we used package OCFit\footnote{https://github.com/pavolgaj/OCFit} \citep{2019OEJV..197...71G}. We found that the 3$^{\rm rd}$ body orbiting eclipsing system with 7370 days period ($\approx$ 20.2 years) on a slightly eccentric orbit. The resulting parameters of its orbit are listed in Tab.~\ref{tab:oc_param}. We did not detect any secular quadratic period changes in the data used in our analysis, which contradicts the findings of \citet{2006PASJ...58..361Z} and \citet{2008PASP..120..720L}.

VY UMi has several CCD minima times published from 1999 and few visual minima times, which were omitted from our analysis. Linear ephemeris determined from a weighted least-squares solution (as in previous case) is:
\begin{equation}
  \mathrm {Min~I} = \mathrm {HJD~} 24552500.0090(7) + 0^{d}. 3254048(4) \times E.  
  \label{eq:lin_vyumi}
\end{equation}
Eclipsing binary GSC 04364-0064 has no published minima times. For our analysis we have used minima times determined from \textit{TESS} observations and 2 our new minima times. A weighted least-squares solution using all minima (weights were calculated as in the previous cases) leads to the linear ephemeris:
\begin{equation}
  \mathrm {Min~I} = \mathrm {HJD~} 24559309.7281(3) + 0^{d}.862851(5) \times E.  
  \label{eq:lin_gsc}
\end{equation}
Ephemerides (\ref{eq:lin_vyumi}) and (\ref{eq:lin_gsc}) were used to create the $O-C$ diagrams of VY UMi and  GSC 04364-0064 as displayed in Fig.~\ref{fig:oc_all}. We can clearly see, that in both cases, quadratic variations are detected, which indicates mass transfer between components and/or magnetic braking. The quadratic ephemerides of both systems are given in Tab.~\ref{tab:oc_param}.

\section{Light curve analysis}
\label{section:lcanalysis}
\begin{table}[t]\centering
  \setlength{\tabnotewidth}{0.99\columnwidth}
  \tablecols{7}
  \setlength{\tabcolsep}{1.\tabcolsep}
  \caption{Photometric parameters of the studied systems } 
\label{tab:parameters}
\small

\setlength{\tabcolsep}{1pt}
\small
\begin{tabular}{l||cc||cc||cc}
\hline
\hline
      & \multicolumn{2}{c||}{\textbf{RU~UMi}}       & \multicolumn{2}{c||}{\textbf{VY~UMi}}    & \multicolumn{2}{c}{\textbf{GSC 04364-00648}} \\
                      & Primary      &  Secondary                    & Primary      &  Secondary                    & Primary      &  Secondary \\ 
                      
\hline                    
$i$ [deg]               &\multicolumn{2}{c||}{$83.4^{+0.15}_{-0.22}$}         &\multicolumn{2}{c||}{$85.1^{+0.22}_{-0.25}$}         &\multicolumn{2}{c}{$76.9^{+0.20}_{-0.21}$}        \\
$q_p$ (M$_{2}$/M$_{1}$)   &\multicolumn{2}{c||}{$0.341^{+0.010}_{-0.006}$}      &\multicolumn{2}{c||}{$0.536^{+0.023}_{-0.025}$}       &\multicolumn{2}{c}{$0.440^{+0.014}_{-0.015}$}      \\
$T$ [K]               &$7420^{a}$     & $4885^{+148}_{-201}$      &$5340^{a}$     &$4850^{+240}_{-250}$         &$7970^{a}$     &$4065^{+44}_{-49}$       \\
$\Omega$              &$2.632^{+0.013}_{-0.007}$&$2.568^{+0.037}_{-0.019}$  &\multicolumn{2}{c||}{$2.821^{+0.034}_{-0.036}$}   &$4.138^{+0.033}_{-0.033}$&$2.767^{+0.032}_{-0.030}$\\
\hline
$l^V/l^V_{tot}$       &$0.93^{+0.02}_{-0.03}$&$ 0.07^{+0.09}_{-0.10}$ &$0.68^{+0.09}_{-0.10}$&$0.32^{+0.06}_{-0.07}$   &$0.92^{+0.05}_{-0.06}$&$0.08^{+0.03}_{-0.03}$\\ 
$\Omega_{crit}$       &\multicolumn{2}{c||}{$2.553^{+0.022}_{-0.012}$} &\multicolumn{2}{c||}{$2.943^{+0.032}_{-0.00.035}$} &\multicolumn{2}{c}{$2.760^{+0.029}_{-0.028}$} \\
$R^{eq}$[SMA]       &$0.457^{+0.001}_{-0.002}$&$ 0.286^{+0.001}_{-0.001}$ &$0.465^{+0.002}_{-0.001}$&$0.357^{+0.001}_{-0.001}$   &$0.273^{+0.002}_{-0.001}$&$0.308^{+0.002}_{-0.002}$\\ 


\hline\hline
\end{tabular}
{\raggedright $^a$ - Temperature of the primary component for all systems were adopted from \citet{2022arXiv220605989B} }
\end{table}

For the analysis of light curves of all three systems, we have relied on the ELISa\footnote{ https://github.com/mikecokina/elisa} code \citep{2021A&A...652A.156C}. It is a cross-platform Python
software package dedicated to modeling close eclipsing binaries including surface features such as spots and pulsations. ELISa utilizes modern approaches to the EB modeling with an emphasis on computational speed while maintaining a sufficient level of precision to process a ground-based and space-based observation. It was designed for easy use even by a not very experienced user. In this paper, we take advantage of its capability to model the light curves of close eclipsing binaries with the builtin capability to solve an inverse problem using least squares thrust region reflective algorithm and Markov Chain Monte-Carlo (MCMC) methods (for references see \citet{2021A&A...652A.156C}).

At the beginning of the fitting process, it is necessary to prepare the input data. ELISa requires phased light curves with normalized flux. All our phased observations in all passbands were transformed to flux and normalized according to flux in the maxima and were simultaneously fitted by the least-square method to find the global optimal solution. 
Subsequently, MCMC sampling was used to produce 1$\sigma$ confidence intervals of the fitted system's parameters. 

Each system was fitted with model containing 5 free parameters: orbital inclination $i$, photometric mass ratio $q_p$, surface potentials of both components $\Omega_1$ and $\Omega_2$ and the effective temperature of the secondary component $T^{eff}_2$. Temperature of the primary component $T^{eff}_1$ for all systems were adopted from \citet{2022arXiv220605989B} and was fixed during fitting process, while temperatures of the secondary component were fitted with no restrictions.

For the components with convective envelopes (effective temperatures bellow $\sim$7000\,K), the albedos $A_1$, $A_2$ of components were set to 0.6 \citep{Rucinski1969}  and gravity darkening factors, $g_1$ and $g_2$ to 0.32 \citep{Lucy1967}. In the case of radiative envelope (above $\sim$7000\,K), the values of albedo and gravity darkening factor were both set to 1.0. \citet{2003IAUS..210P.A20C} models of stellar atmospheres were used. The linear limb darkening coefficients for each component were interpolated from the \citet{vanHamme1993} tables. 

The weights of individual data points were established as $1/\sigma^2$, where $\sigma$ is the standard error of point derived during photometric measurement. Initially, the least-square algorithm was used with suitable initial parameters to find an approximate solution and then the parameter space near the solution was explored with MCMC sampler with 500 walkers and 500 iterations with prior 300 iterations discarded as it belonged to the thermalization stage of the sampling. 

\begin{figure}[t]
	\centering
	\includegraphics[width=0.48\columnwidth, angle=0]{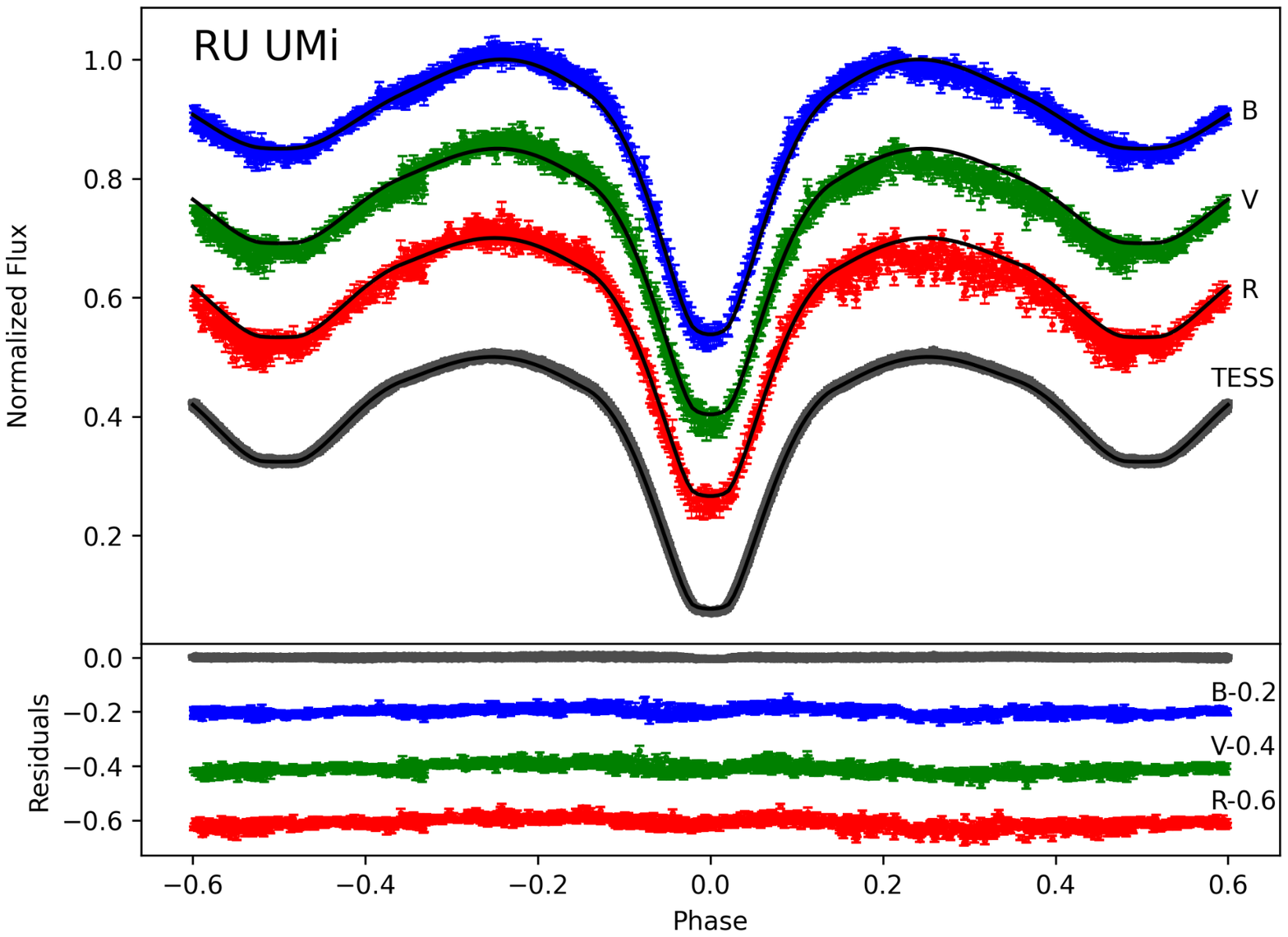}
	\includegraphics[width=0.45\columnwidth, angle=0, trim={4cm 0 0 0},clip]{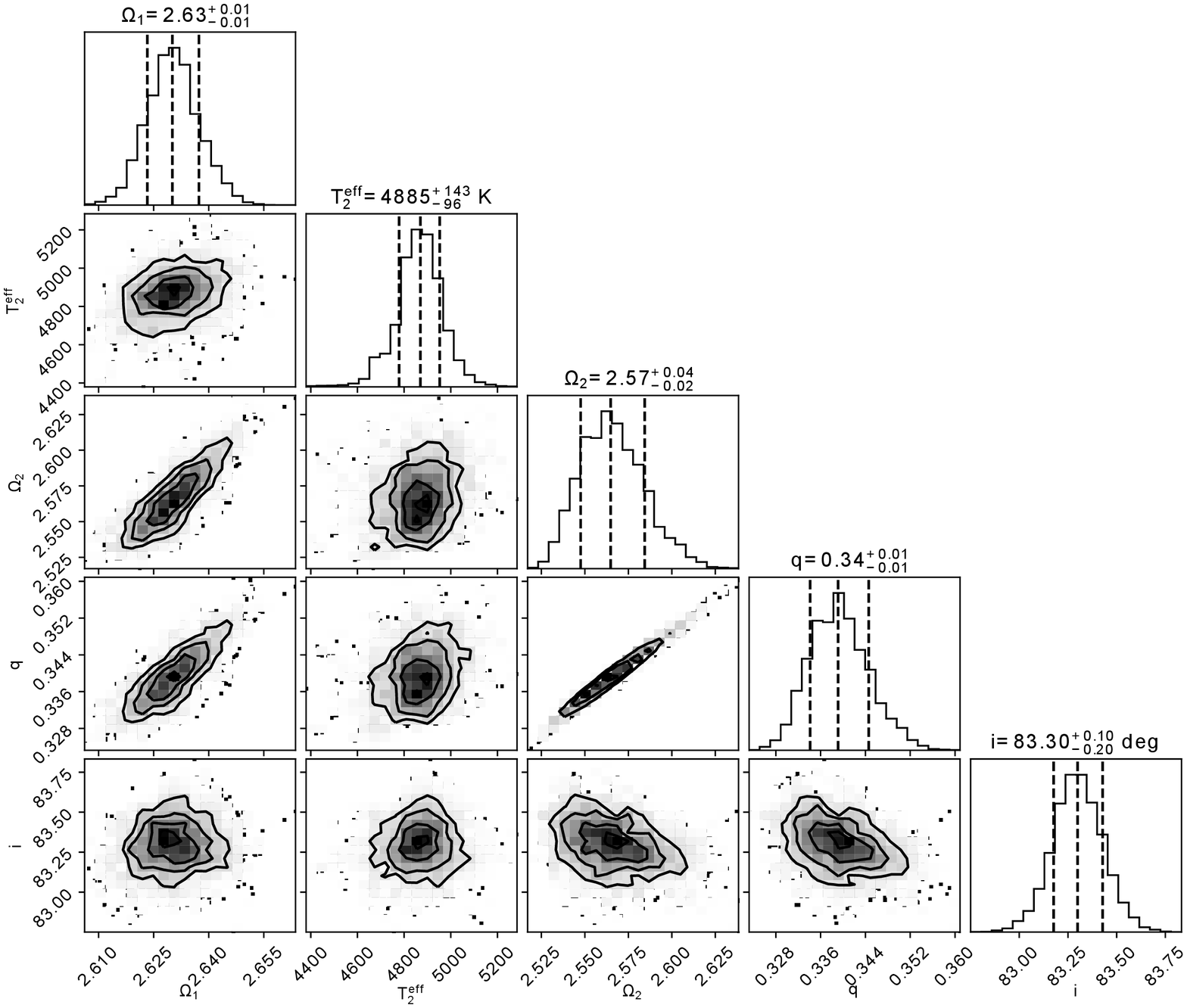}
	\includegraphics[width=0.48\columnwidth, angle=0]{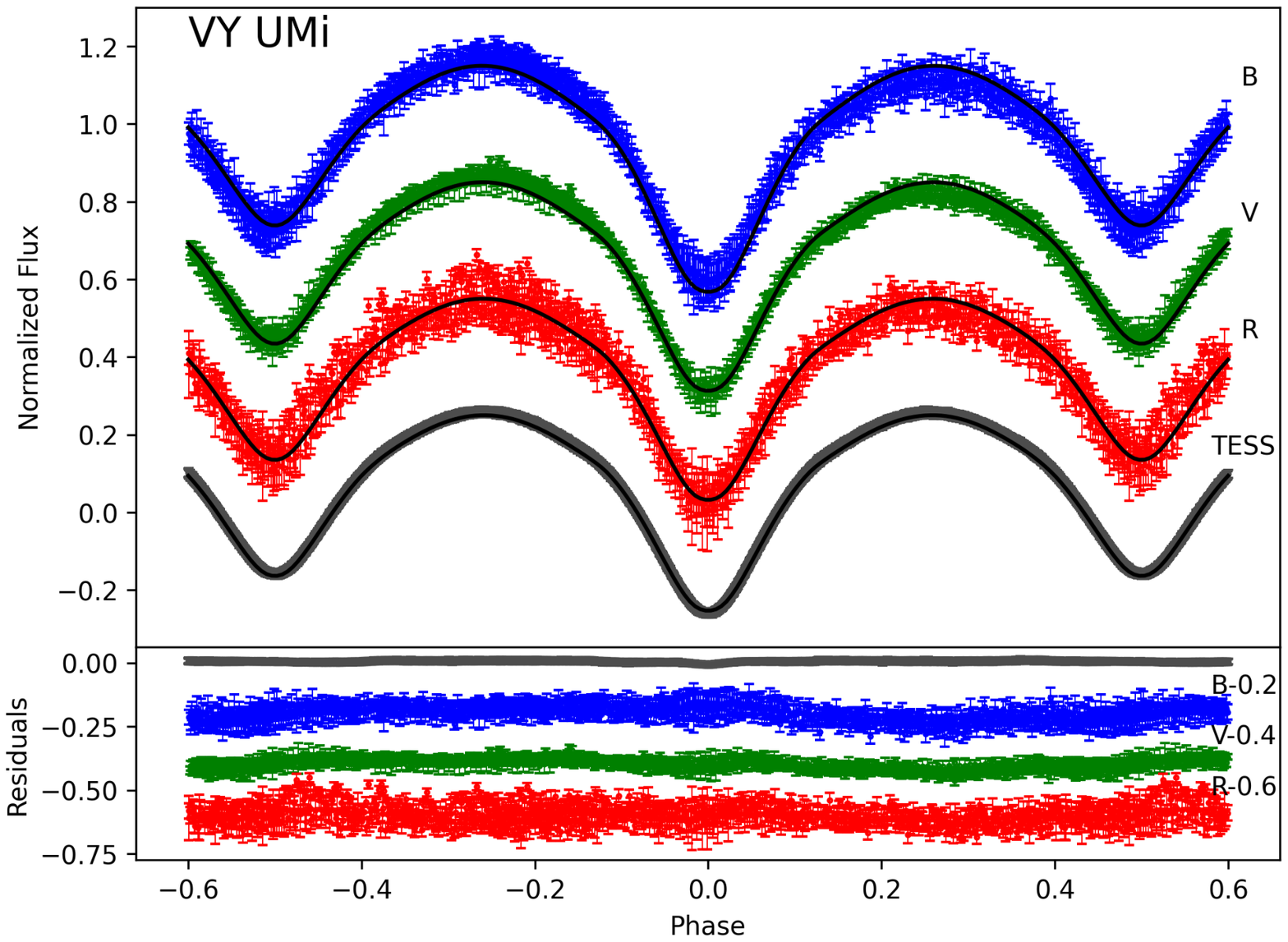}
	\includegraphics[width=0.43\columnwidth, angle=0, trim={5cm 0 0 0},clip]{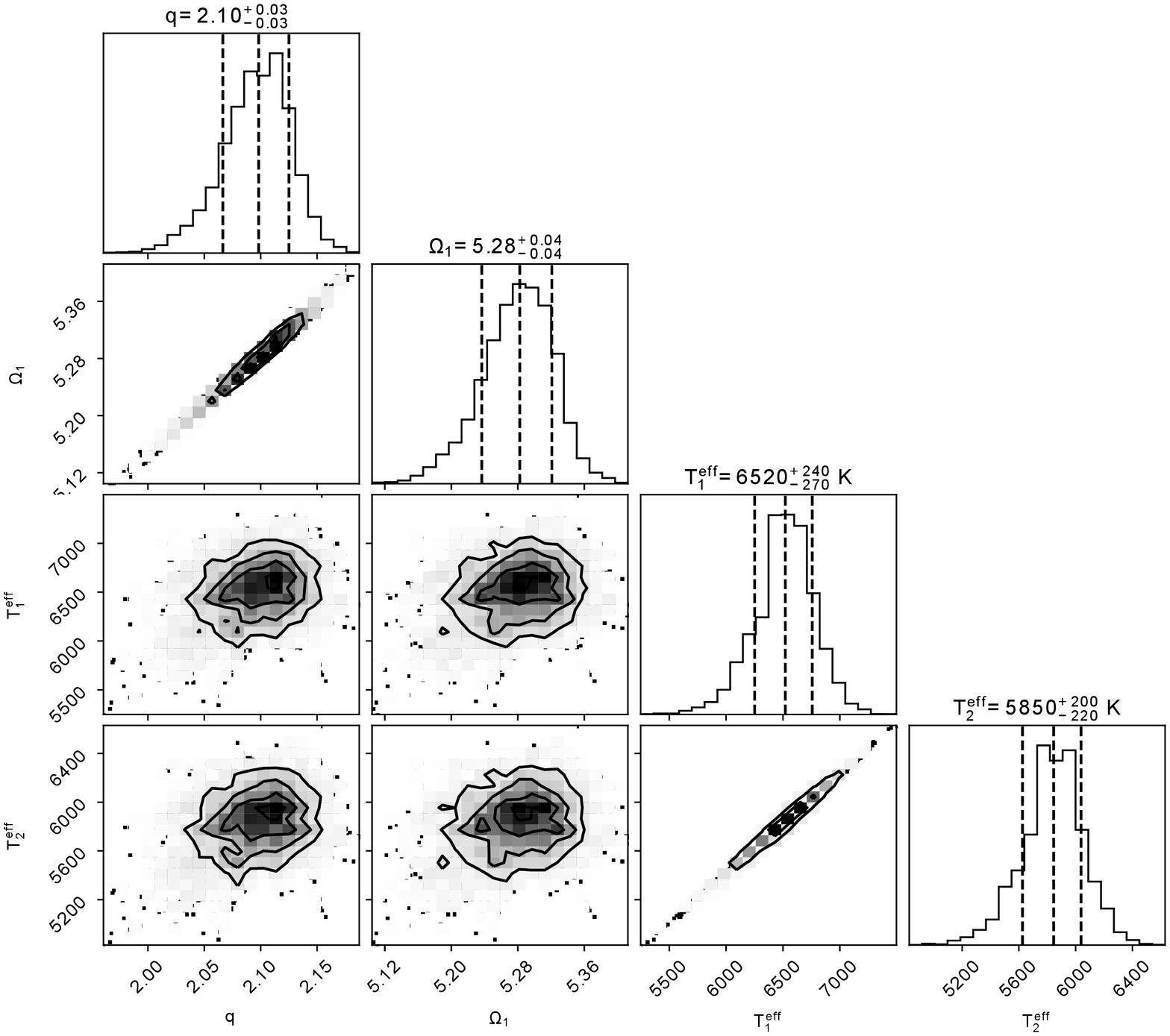}
	\includegraphics[width=0.48\columnwidth, angle=0]{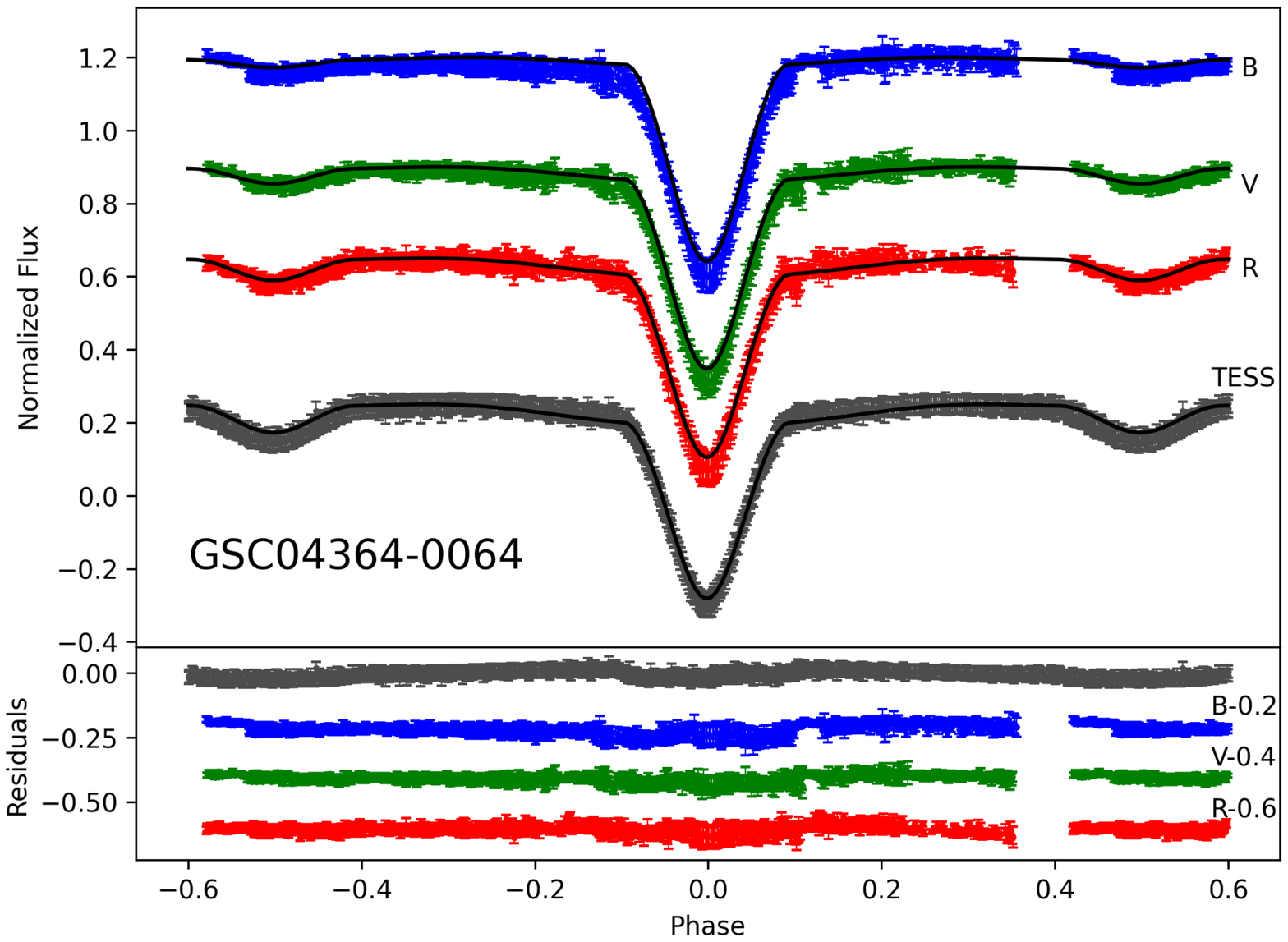}
	\includegraphics[width=0.42\columnwidth, angle=0, 
	                            trim={4cm 0 0 0},clip]{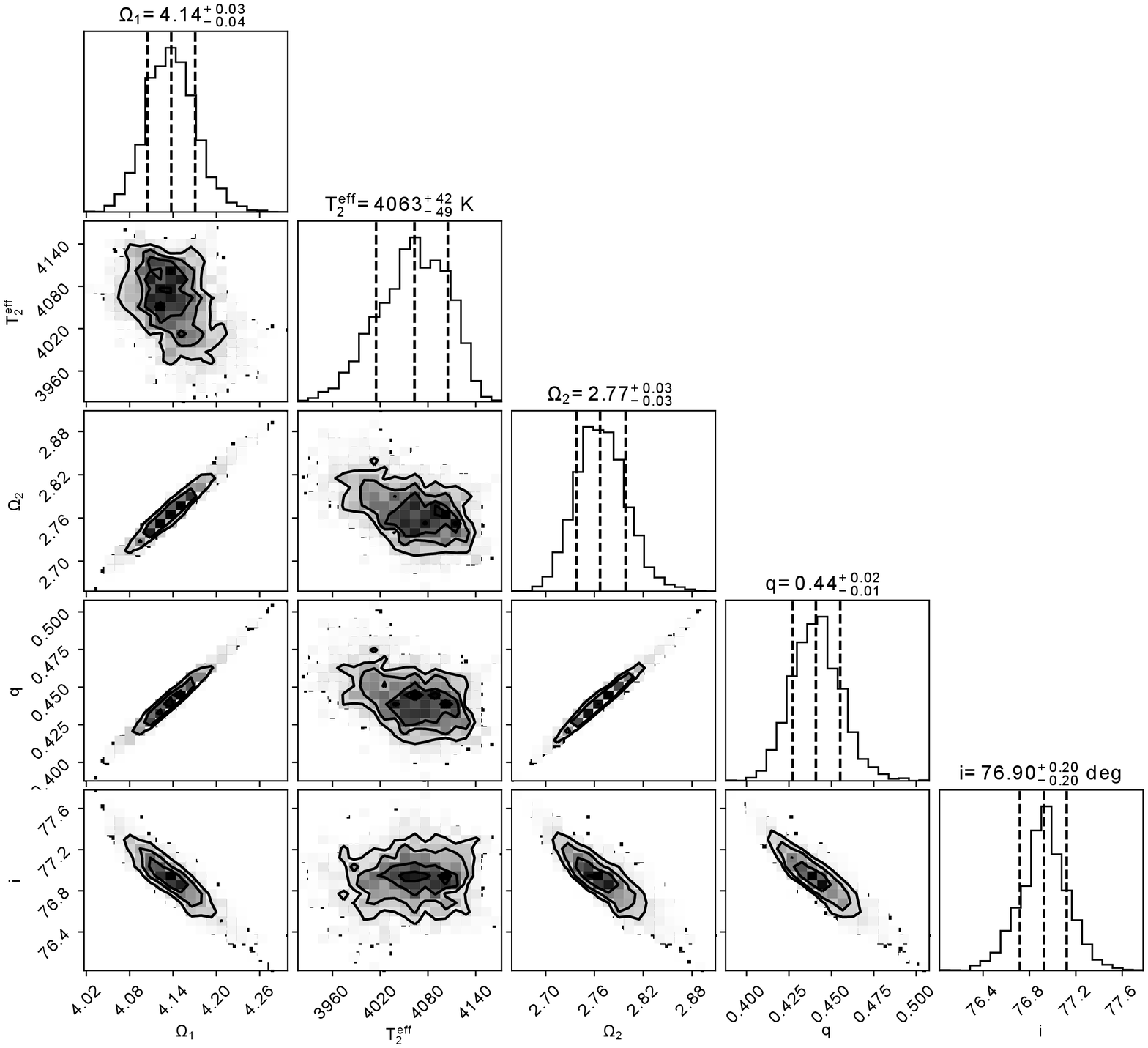}
	\caption{The synthetic model fitted on observational data of (from top) RU~UMi, VY~UMi and GSC 04364-0064 together with the corresponding results of the MCMC sampling displayed in the form of the corner plot.}
	\label{fig:fit_all}
\end{figure}

\begin{figure}[!ht]
	\centering
    \includegraphics[width=0.54\columnwidth]{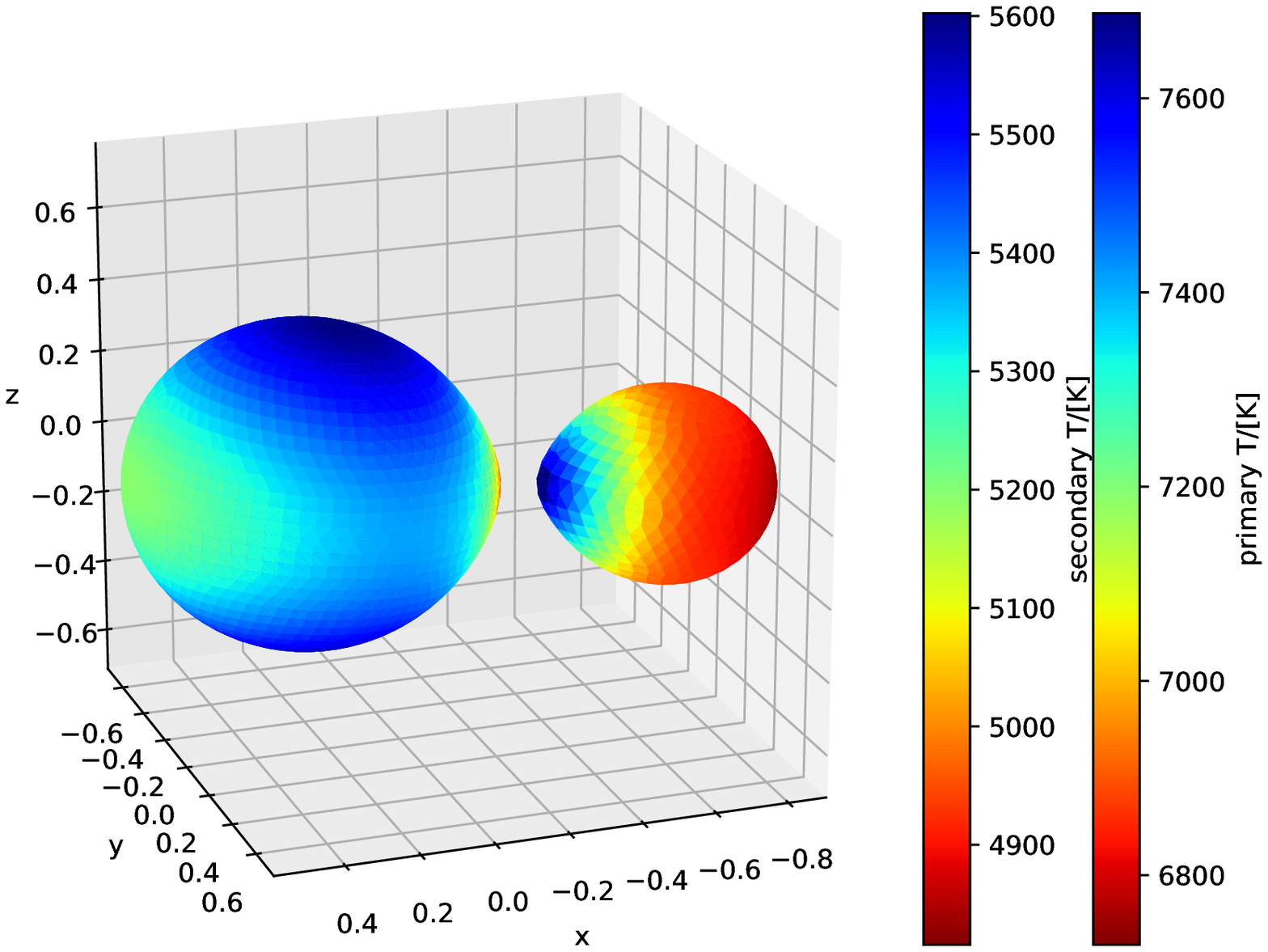}
    \includegraphics[width=0.50\columnwidth]{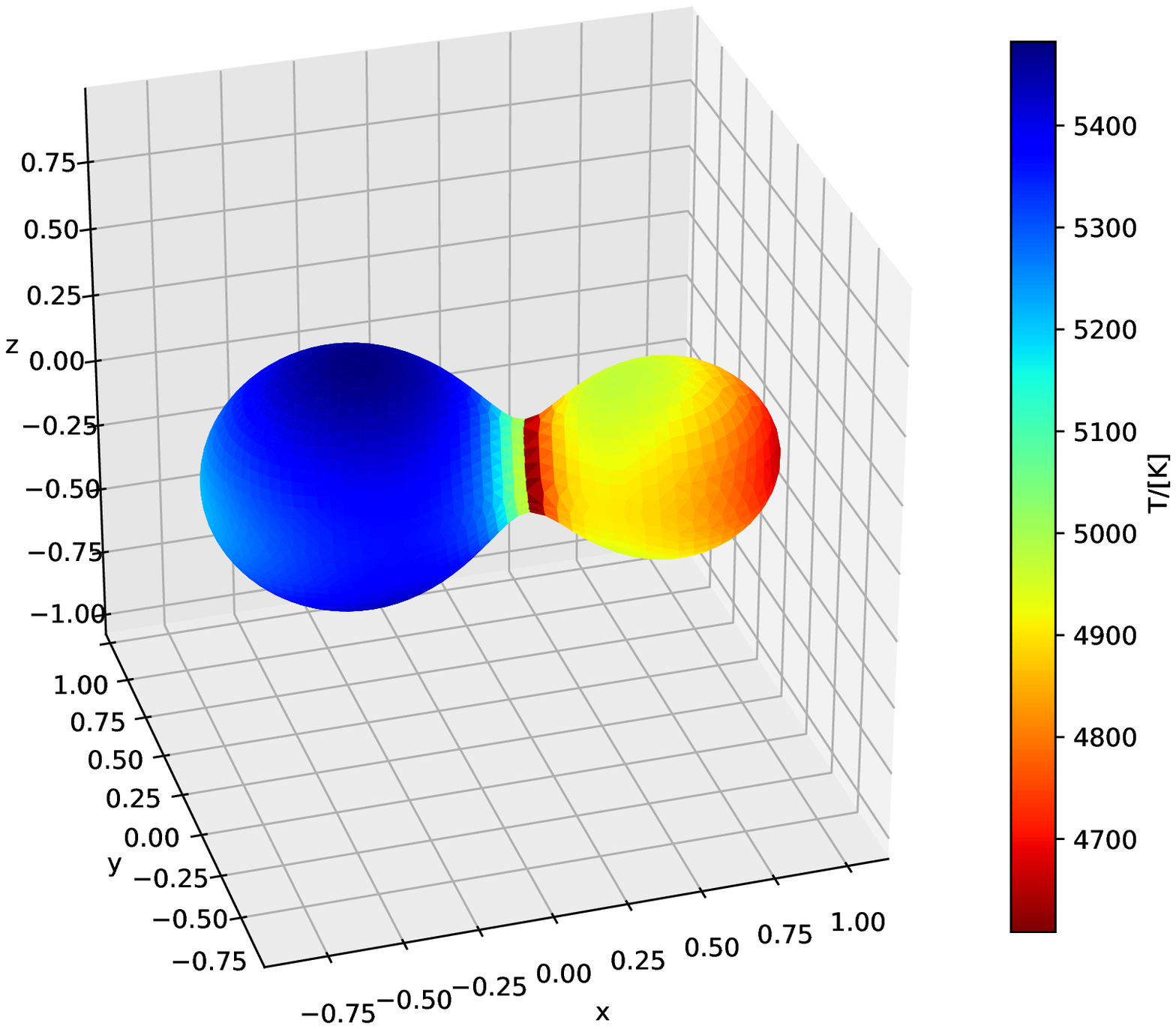}
    \includegraphics[width=0.54\columnwidth]{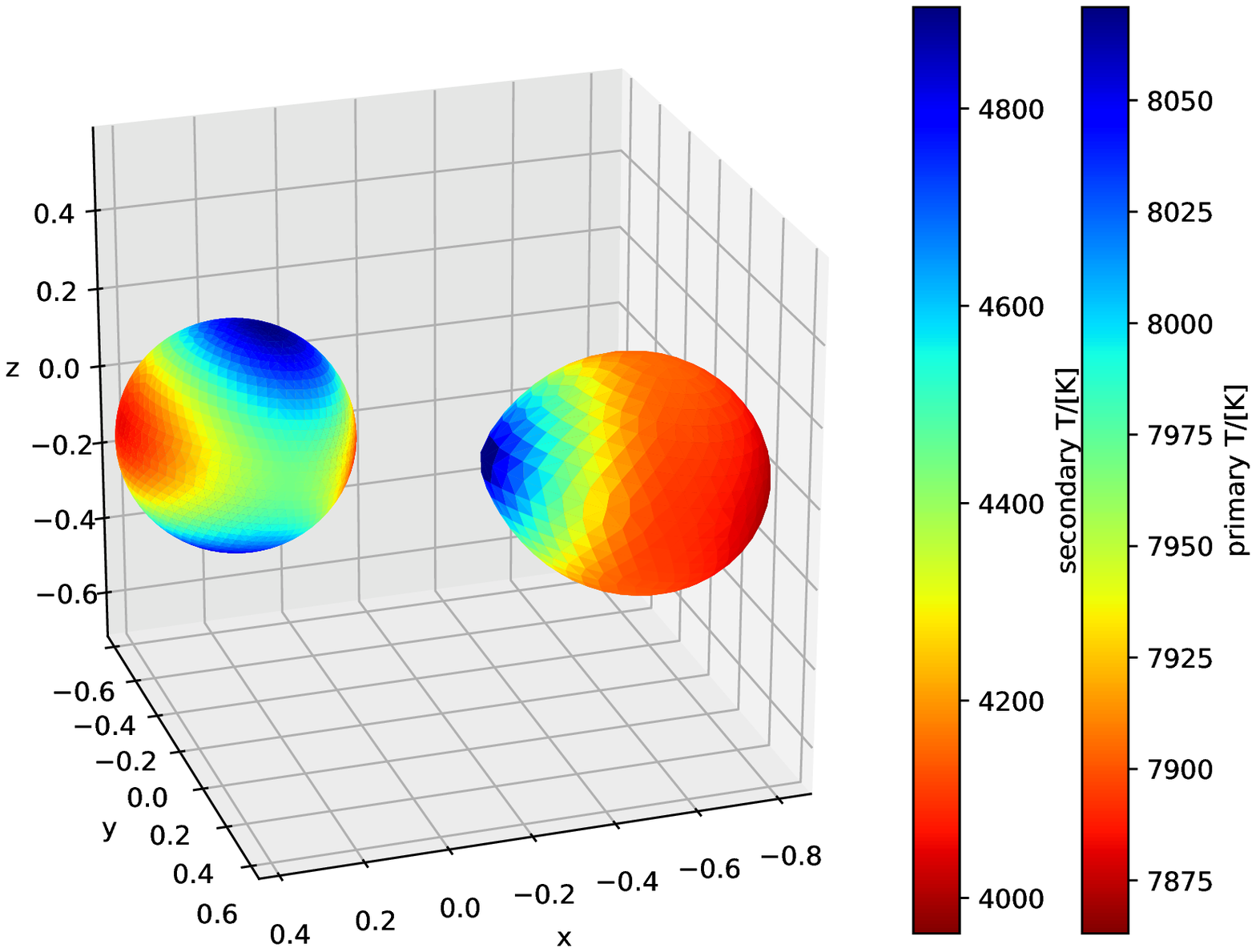}

	\caption{3D models with the surface temperature distributions of (from top) RU~UMi, VY~UMi and GSC 04364-0064 }
	\label{fig:3d_all}
\end{figure}

The resulting as well as derived parameters of all systems, like relative luminosities of the components in $V$ filter  $l^V_{1,2}/l_{tot}^V$, a critical potential $\Omega_{crit}$, corresponding equivalent radius $R^{eq}$ in SMA units (semi-major axis) are listed in Tab.~\ref{tab:parameters}. The best-fit models with observed LCs and resulting flat chains displayed in the form of the corner plot are shown in Fig.~\ref{fig:fit_all} and 3D models with the surface temperature distributions are shown in 
Fig.~\ref{fig:3d_all}.

\section{Absolute parameters of the systems}
\label{sec:abs_param}
The absolute parameters of the binary components, like their masses $M_{1,2}$, radii $R_{1,2}$, luminosities $L_{1,2}$ and  semi-major axis of the orbit $a$, can be mainly determined by the combination of photometric solution and analysis of radial velocity curve. Radial velocities are available only for RU UMi system \citep{1988PASJ...40...79O, 1996Obs...116..288M}. We used their measurements and re-analyze them with ELISa code (assuming circular orbit) and determined orbital and absolute parameters of the components as listed in Tab.\ref{tab:abs_parameters_ruumi}. 

\begin{table}[t]\centering
  \setlength{\tabnotewidth}{0.99\columnwidth}
  \tablecols{4}
  \setlength{\tabcolsep}{1.\tabcolsep}
  \caption{Parameters of the spectroscopic orbit and absolute parameters of RU UMi system derived from radial velocity and light curve solutions.} 
\label{tab:abs_parameters_ruumi} 
\small
\setlength{\tabcolsep}{1pt}
\small
\begin{tabular}{l|c||r|c}
\hline
\hline
                      
\hline 
$M_1$ [$M_\odot$] &$2.65^{+0.11}_{-0.16}$     &$K_1$ [$km/s$]&$96.5^{+6.4}_{-5.9}$            \\
$M_2$ [$M_\odot$]  &$0.85^{+0.12}_{-0.10}$    &$K_1$ [$km/s$]&$301.6^{+10.8}_{-8.3}$         \\
$R_1$ [$R_\odot$]& $1.89^{+0.7}_{-0.4}$       &$q$&$0.32^{+0.02}_{-0.02}$ \\
$R_2$ [$R_\odot$] &$1.18^{+0.4}_{-0.3}$       & $a \sin i$ [$R_\odot$] &$4.13^{+0.11}_{-0.08}$ \\
$L_1$ [$L_\odot$] & $9.61^{+2.51}_{-1.48}$      & $\gamma$ [$km/s$] &$-21.0^{+3.8}_{-4.0}$       \\
$L_2$ [$L_\odot$] & $0.71^{+0.16}_{-0.12}$    &                  &                           \\
$a$ [$R_\odot$] & $4.15^{+0.12}_{-0.09}$      &                 & \\

\hline                                                                                              
\hline
\end{tabular}
\end{table}

But if we know the distance to object from independent measurement (e.g parallaxes from \textit{GAIA} measurements), we can find absolute parameters from properties of binary derived from photometric solution and using basic relationships, described in the following text.

Let us assume that we have calculated $q_p$, $i$, $T_{1,2}$, $R^{eq}_{1,2}$ and $l^V_{1,2}/l_{tot}^V$ from analysis of the light curves and we know standard $V$ magnitude of system in phase 0.25. 
Absolute magnitude $M_V$ of the system we can find from equation
\begin{equation}
  M^V = V - 5\log(d) +5 - A^V,  
  \label{eq:distance_modulus}
\end{equation}
where $d$ is distance and extinction coefficient $A^V$ can be determined from the dust map in \citet{2019ApJ...887...93G}. The absolute magnitudes of each component can be find from relation
\begin{equation}
  M^V_{1,2} - M^V = -2.5\log \frac{l^V_{1,2}}{l_{tot}^V}.  
  \label{eq:abs_mag}
\end{equation}
Corresponding bolometric magnitude is 
\begin{equation}
  M^{Bol}_{1,2} = M^V_{1,2}+BC,  
  \label{eq:bol_mag}
\end{equation}
where bolometric corection is from \citet{2020MNRAS.496.3887E}. If we assume that bolometric magnitude of the Sun is $M^{Bol}_{\odot}$=4.73 mag \citep{2010AJ....140.1158T}, the luminosities of components can be found from 
\begin{equation}
  M^{Bol}_{1,2} - M^{Bol}_{\odot} = -2.5\log \frac{L_{1,2}}{L_\odot}, 
  \label{eq:luminosities}
\end{equation}
and corresponding radii
\begin{equation}
  R_{1,2}=\sqrt{\frac{L_{1,2}}{4\pi\sigma T_{1,2}^4}}.
  \label{eq:radii}
\end{equation}
The distance between components can be found using their equivalent radii $R^{eq}_{1,2}$ 
\begin{equation}
  a=\frac{1}{2}\left( \frac{R_{1}}{R^{eq}_{1}} + \frac{R_{2}}{R^{eq}_{2}} \right).
  \label{eq:distance}
\end{equation}
\begin{table}[t]\centering
  \setlength{\tabnotewidth}{0.99\columnwidth}
  \tablecols{7}
  \setlength{\tabcolsep}{1.\tabcolsep}
  \caption{Absolute parameters of the studied systems derived from their distances and photometric solutions.} 
\label{tab:abs_parameters} 
\small

\setlength{\tabcolsep}{1pt}
\small
\begin{tabular}{l||cc||cc||cc}
\hline
\hline
   & \multicolumn{2}{c||}{\textbf{RU~UMi}}       & \multicolumn{2}{c||}{\textbf{VY~UMi}}    & \multicolumn{2}{c}{\textbf{GSC 04364-00648}} \\
                      & Primary      &  Secondary                    & Primary      &  Secondary                    & Primary      &  Secondary \\ 
                      
\hline                    
$M$ [$M_\odot$]    &1.90(21)       & 0.64(17)     &0.97(15)     &0.52(18)     &2.54(42)     &1.12(31)            \\
$R$ [$R_\odot$]   & 1.55(11)       & 1.16(9)         &0.96(2)     &0.87(2)     &1.22(3)   &2.25(5)            \\
$L$ [$L_\odot$]   &6.61(1.47)      & 0.69(14)        &0.67(10)     &0.38(11)     &5.40(1.34)   &1.24(23)       \\
$a$ [$R_\odot$]   &\multicolumn{2}{c||}{3.74(37)} &\multicolumn{2}{c||}{2.27(21)} &\multicolumn{2}{c}{5.88(79)}        \\

$A^V$ [mag]        &\multicolumn{2}{c||}{0.0} &\multicolumn{2}{c||}{0.0} &\multicolumn{2}{c}{0.03(3)}    \\  
$M^{Bol}$ [mag] &2.67(7)     & 5.12(13)  & 5.16(8)   &    5.75(14)     & 2.89(8) & 4.49(11)       \\
$BC$ [mag]      &0.06        & -0.31    & -0.082     &  -0.30         & 0.03    & -1.03          \\
$d$ [pc]        &\multicolumn{2}{c||}{283.0(1.2)} &\multicolumn{2}{c||}{164.5(3)} &\multicolumn{2}{c}{512.5(4.8)}    \\  

\hline                                                                                              
\hline
\end{tabular}
\end{table}
Total mass $M_1+M_2$ of the system we can derive using Kepler's 3$^{\rm rd}$ law 
\begin{equation}
  \frac{a^3}{P^2} = \frac{G(M_1+M_2)}{4\pi^2},
  \label{eq:mass}
\end{equation}
and individual masses can be found from mass ratio $q_p$.

The absolute parameters for the studied objects determined by the method described above are listed in Tab.~\ref{tab:abs_parameters}.  The uncertainties of the parameters were calculated considering the errors  of the light curve solutions of the systems and errors in their distances. 

\section{DISCUSSION AND CONCLUSIONS}
In our study, we have presented the photometric analysis of multi-color $BVR$ and \textit{TESS} photometry of three eclipsing binaries, RU~UMi, VY~UMi and GSC 04364-00648, for the first time for the last two systems. We have also analyzed their period variations considering archival data and our new minima times.
The presented photometry solutions, mainly for VY~UMi and GSC 04364-00648, has some small disagreements with observations, the residuals at some phases show up to 0.1 deviations in normalized flux value. These issues can be caused by spots, weather conditions, etc. They are really small, and we can not even try to explain what phenomena they are caused by. Future observation is needed for this purpose.

RU~UMi has been studied in the past years by several authors. Recently, \citet{2008PASP..120..720L} analyzed period variation, fitted light-curve and determined absolute parameters of the components from radial velocities solution.
From their period analysis authors concluded that long-term period changes can be caused by the combination of angular momentum loss (AML) and mass transfer from the less massive secondary to the more massive primary. In our period analysis, we used only minima times obtained from ground-based photoelectric and CCD observations as well as satellite observations from \textit{TESS}, where we can expect higher precision with respect to older visual and photographic observations. We detected wave-like variations with low-amplitude ($\sim 5$ minutes) in $O-C$ residua. It can be interpreted as a consequence of the light-time effect caused by the 3$^{\rm rd}$ invisible component. From the parameters listed in Tab.~\ref{tab:oc_param} we can see that the orbital period of the 3$^{\rm rd}$ body is 7370 days and the orbit is slightly eccentric. According to the mass function of the 3$^{\rm rd}$ body $f(m_3)$ and masses of the binary components (see Tab.~\ref{tab:abs_parameters}), we can find that the minimum mass of the 3$^{\rm rd}$ component in the case of the edge-on orbit ($\sin i_3 = 1$) should be $M_3$=0.063(16)$M_\odot \sim 60 M_J$. It corresponds to a low massive red dwarf or more probably (due to its mass) it is a brown dwarf \citep{2014ASSL..401.....J} with very low luminosity. It is supported also by the results of the photometric solution, where no 3$^{\rm rd}$ light was detected. Photometric analysis of $BVR$ and \textit{TESS} light curves confirmed previous findings that RU~UMi is a near contact system with a secondary component that almost fulfills its Roche lobe. We were not able to find any satisfactory LC solution with spot(s) (not even\textit{TESS} LC) as was done by \citet{2008PASP..120..720L}, although some wave-like variation is visible in residuals in Fig.~\ref{fig:fit_all}. We can explain it by the temporal evolution of the spot, when spot's parameters such as diameter, temperature and position on the star's surface are changing during decades. 
Our absolute parameters of components determined from radial velocity solution are little bigger than presented in  \citet{2008PASP..120..720L}. On the other side, absolute parameters determined from \textit{GAIA} parallax are smaller than previously determined and correspond to A6V primary component and evolved K5 star secondary. This differences deserves deeper analysis, because many factors can affect results. One of them is a mass ratio which has strong influence to partial dimensions of the components as well as inclination due to $q-i$ correlation. \citet{2005Ap&SS.296..221T} showed that the photometric mass ratio for semi-detached and over-contact binaries is often overestimated for partial eclipses. Recently \citet{2022Galax..10....8T} noted that not properly modeled third light will lead to mass ratios that are too low. Our solution of RU~UMi show no third light and no total eclipses on the LCs (just like other stars). Presented photometric mass ratios $q_p$ have to be considered as a high estimates and this affects also determination of absolute parameters from distances. 

Photometric analysis of VY~UMi showed that the system is a typical W~UMa type overcontact binary with a more massive primary component ($q_p=0.535$). Its orbital period and determined temperatures of both components place the system in a W-type subclass of overcontact binaries. The detected parabolic period change reflected on the $O-C$ diagram can be explained by the mass transfer from a less massive star to a more massive one. Period increase with rate 2.56(9)$\times10^{-7}$~d/yr$^{-1}$ detected in the VY~UMi system corresponds to mass transfer from the secondary to the primary component.

The first photometric solution of GSC 04364-00648 light curves revealed that the system is semi-detached binary, where a cool secondary component almost fills its Roche lobe as detected in some other near-contact systems, like EG Cep \citep{2013AJ....145...80D} or CR~Tau \citep{2021RAA....21..174K}. Although we can see some quadratic changes on the $O-C$ diagram, which corresponds to a period decrease with a high rate of $-2.26(5)\times10^{-5}$~d/yr$^{-1}$, we cannot make strict conclusions about period variation in the system, mainly due to short time (2019-2021) and uneven coverage of $O-C$ diagram. We have to wait for other observations to confirm or disprove this trend.

\acknowledgments
\section*{Acknowledgments}
This work was supported by Ukrainian national grant 0122U000937 and by the Slovak Research and Development Agency under contract No. APVV-20-0148. The research of P.G. was supported by the internal grant No. VVGS-PF-2021-2087 of the Faculty of Science, P. J. {\v S}af{\'a}rik University in Ko{\v s}ice.

\bibliography{kudak.bib}

\newpage
\clearpage
\begin{appendices}

\small
\tiny
\begin{longtable}{|l|l|l|l|l|l|l|}
\caption{RU UMi times of minima determined from TESS lightcurves. Errors are in parenthesis.} 
\label{tab:tmin_RU_UMi} \\
\hline 

\textbf{BJD} & 
\textbf{BJD} & 
\textbf{BJD} & 
\textbf{BJD} & 
\textbf{BJD} &
\textbf{BJD} &
\textbf{BJD}\\
\hline 

\endfirsthead

\multicolumn{7}{c}%
{{\bfseries \tablename\ \thetable{} -- continued from previous page}} \\
\hline 
\textbf{BJD} & 
\textbf{BJD} & 
\textbf{BJD} & 
\textbf{BJD} & 
\textbf{BJD} &
\textbf{BJD} &
\textbf{BJD}\\
\hline 
\endhead

\hline \multicolumn{7}{|r|}{{Continued on next page}} \\ \hline
\endfoot

\hline \hline
\endlastfoot

58683.4591(5)   & 58706.2924(2)  & 58730.7012(1)  & 58886.8666(2)  &  58910.7460(11) & 59428.8517(1)  & 59585.0166(2)   \\   
58683.7210(10)  & 58706.5541(1)  & 58730.9632(2)  & 58887.1296(1)  &  58911.0137(7)  & 59429.1138(2)  & 59585.5416(2)   \\  
58683.9827(1)   & 58706.8163(2)  & 58731.2262(9)  & 58887.3914(2)  &  58911.2710(10) & 59429.3768(1)  & 59585.8049(2)   \\  
58684.2447(2)   & 58707.0794(1)  & 58731.4882(2)  & 58887.6544(1)  &  58911.5387(7)  & 59429.6388(2)  & 59586.0665(2)   \\  
58684.5074(1)   & 58707.3413(2)  & 58731.7511(1)  & 58887.9164(2)  &  58911.7950(20) & 59429.9018(1)  & 59586.3312(2)   \\  
58684.7694(2)   & 58707.6043(1)  & 58732.0130(21) & 58888.1793(1)  &  58913.9016(3)  & 59430.1637(2)  & 59586.5911(2)   \\  
58685.0323(1)   & 58707.8672(2)  & 58732.2761(1)  & 58888.4413(2)  &  58914.1627(2)  & 59430.4267(1)  & 59586.8546(2)   \\  
58685.2944(2)   & 58708.1292(1)  & 58732.5380(20) & 58888.7046(1)  &  58914.4261(2)  & 59430.6886(2)  & 59587.1164(2)   \\  
58685.5573(1)   & 58708.3912(2)  & 58732.8008(1)  & 58888.9664(2)  &  58914.6877(2)  & 59430.9515(1)  & 59587.3796(2)   \\  
58685.8194(2)   & 58708.6541(1)  & 58733.0630(21) & 58889.2293(1)  &  58914.9513(3)  & 59431.2135(2)  & 59587.6413(2)   \\  
58686.0823(1)   & 58708.9161(2)  & 58733.3253(2)  & 58889.4912(2)  &  58915.2128(2)  & 59431.4764(1)  & 59587.9044(2)   \\  
58686.3443(2)   & 58709.1791(1)  & 58733.5904(2)  & 58889.7543(1)  &  58915.4765(3)  & 59431.7385(2)  & 59588.1662(2)   \\  
58686.6071(1)   & 58709.4411(2)  & 58733.8509(1)  & 58890.0161(2)  &  58915.7377(1)  & 59432.0014(1)  & 59588.4297(2)   \\  
58686.8692(2)   & 58709.7040(1)  & 58734.1127(2)  & 58890.2793(1)  &  58916.0014(3)  & 59432.2631(1)  & 59588.6911(2)   \\  
58687.1321(1)   & 58709.9671(2)  & 58734.3757(1)  & 58890.5410(21) &  58916.2626(1)  & 59432.5255(6)  & 59588.9543(2)   \\  
58687.3941(2)   & 58711.8037(1)  & 58734.6377(2)  & 58890.8042(1)  &  58916.5259(4)  & 59433.8383(2)  & 59589.2161(2)   \\  
58687.6569(1)   & 58712.0657(2)  & 58734.9006(1)  & 58891.0661(2)  &  58916.7789(8)  & 59434.1010(10) & 59589.4791(2)   \\  
58688.1820(12)  & 58712.3286(1)  & 58735.1626(2)  & 58891.3293(1)  &  58917.0450(9)  & 59434.3631(2)  & 59589.7410(20)  \\  
58688.4440(23)  & 58712.5906(2)  & 58735.4257(1)  & 58891.5913(1)  &  58917.3119(7)  & 59434.6260(10) & 59590.0042(2)   \\  
58688.7068(1)   & 58712.8536(1)  & 58735.6874(2)  & 58891.8539(1)  &  58917.5720(11) & 59434.8880(21) & 59590.2660(20)  \\  
58688.9690(21)  & 58713.1156(2)  & 58735.9504(1)  & 58892.1159(2)  &  58917.8376(6)  & 59435.1510(9)  & 59590.5291(2)   \\  
58689.2316(1)   & 58713.3785(1)  & 58736.2125(2)  & 58892.3788(1)  &  58918.0950(21) & 59435.4130(22) & 59590.7908(2)   \\  
58689.4937(2)   & 58713.6405(2)  & 58736.4755(1)  & 58892.6407(2)  &  58918.3627(7)  & 59435.6760(10) & 59591.0541(2)   \\  
58689.7567(1)   & 58713.9036(1)  & 58736.7374(2)  & 58892.9037(1)  &  58918.8876(7)  & 59435.9379(2)  & 59591.3158(2)   \\  
58690.0188(2)   & 58714.1654(2)  & 58737.0004(1)  & 58893.1657(2)  &  58919.1440(20) & 59436.2008(1)  & 59591.5790(20)  \\  
58690.2813(1)   & 58714.4286(1)  & 58870.8572(3)  & 58893.4289(1)  &  58919.4125(7)  & 59436.4628(2)  & 59591.8407(2)   \\  
58690.5437(2)   & 58714.6905(2)  & 58871.1191(2)  & 58893.6906(2)  &  58920.2051(9)  & 59436.7257(1)  & 59592.1039(2)   \\  
58690.8065(1)   & 58714.9533(1)  & 58871.3818(1)  & 58893.9536(1)  &  58920.7190(20) & 59436.9877(2)  & 59592.3656(2)   \\  
58691.0687(2)   & 58715.2165(2)  & 58871.6437(2)  & 58894.2156(2)  &  58920.9868(7)  & 59437.2507(1)  & 59592.6290(20)  \\  
58691.3313(1)   & 58715.4772(3)  & 58871.9065(1)  & 58894.4786(1)  &  58921.2450(21) & 59437.5127(2)  & 59594.2037(2)   \\  
58691.5935(2)   & 58716.0030(10) & 58872.1686(2)  & 58894.7404(2)  &  58921.5123(7)  & 59437.7756(1)  & 59594.4653(2)   \\  
58691.8563(1)   & 58716.2651(2)  & 58872.4315(1)  & 58895.0033(1)  &  58921.7690(20) & 59438.0377(2)  & 59594.7287(2)   \\  
58692.1184(2)   & 58716.5281(1)  & 58872.6935(2)  & 58895.2654(2)  &  58922.0371(7)  & 59438.3006(1)  & 59594.9902(2)   \\  
58692.3850(20)  & 58716.7901(2)  & 58872.9569(2)  & 58895.5285(1)  &  58922.2950(21) & 59438.5625(2)  & 59595.2537(2)   \\  
58692.6441(2)   & 58717.0529(1)  & 58873.2186(2)  & 58895.7904(2)  &  58922.5621(7)  & 59438.8254(1)  & 59595.5152(2)   \\  
58692.9063(1)   & 58717.3150(24) & 58873.4816(1)  & 58896.0534(1)  &  58922.8190(20) & 59439.0874(2)  & 59595.7789(2)   \\  
58693.1684(2)   & 58717.5779(1)  & 58873.7435(2)  & 58896.3154(1)  &  58923.0864(7)  & 59439.3504(1)  & 59596.0401(2)   \\  
58693.4313(1)   & 58717.8400(21) & 58874.0064(1)  & 58896.5783(1)  &  58923.3430(22) & 59439.6124(2)  & 59596.3035(2)   \\  
58693.6934(2)   & 58718.1031(1)  & 58874.2683(2)  & 58896.8402(2)  &  58923.8690(20) & 59439.8754(1)  & 59596.5651(2)   \\  
58693.9559(1)   & 58718.3648(2)  & 58874.5313(1)  & 58897.1032(1)  &  58924.1364(7)  & 59440.1374(2)  & 59596.8285(1)   \\  
58694.2182(2)   & 58718.6278(1)  & 58874.7932(2)  & 58897.3653(2)  &  58924.3940(21) & 59440.4003(1)  & 59597.0899(2)   \\  
58694.4807(1)   & 58718.8898(2)  & 58875.0564(1)  & 58897.6277(1)  &  58924.6619(7)  & 59440.6622(2)  & 59597.3534(2)   \\  
58694.7431(2)   & 58719.1527(1)  & 58875.3182(2)  & 58899.4663(1)  &  58924.9190(20) & 59440.9250(10) & 59597.6149(2)   \\  
58695.0062(1)   & 58719.4159(2)  & 58875.5813(1)  & 58899.7285(3)  &  58925.1867(7)  & 59441.1872(2)  & 59597.8782(2)   \\  
58695.2681(2)   & 58719.6771(2)  & 58875.8430(20) & 58899.9897(2)  &  58925.4440(21) & 59441.7120(20) & 59598.1398(2)   \\  
58695.5308(1)   & 58719.9590(20) & 58876.1061(1)  & 58900.2535(3)  &  58925.9690(20) & 59441.9750(10) & 59598.4031(2)   \\  
58695.7930(22)  & 58720.2025(1)  & 58876.3679(1)  & 58900.5156(2)  &  58926.2364(8)  & 59442.2370(20) & 59598.6647(2)   \\  
58696.0559(1)   & 58720.4646(2)  & 58876.6317(6)  & 58900.7786(3)  &  58926.4860(10) & 59442.7619(2)  & 59598.9282(2)   \\  
58697.6307(1)   & 58720.7277(1)  & 58876.8921(3)  & 58901.0399(1)  &  59420.1901(2)  & 59443.0250(10) & 59599.1897(2)   \\  
58697.8928(2)   & 58720.9896(2)  & 58877.1560(12) & 58901.3032(3)  &  59420.4530(10) & 59443.2868(2)  & 59599.4528(2)   \\  
58698.1556(9)   & 58721.2525(1)  & 58877.4178(2)  & 58901.5648(1)  &  59420.7150(20) & 59443.5498(1)  & 59599.7146(2)   \\  
58698.4177(2)   & 58721.5145(2)  & 58877.6809(1)  & 58901.8268(4)  &  59420.9779(9)  & 59443.8117(2)  & 59599.9780(20)  \\  
58698.6805(1)   & 58721.7773(1)  & 58877.9427(2)  & 58902.0899(1)  &  59421.2399(2)  & 59444.0746(1)  & 59600.2408(3)   \\  
58698.9425(2)   & 58722.0394(2)  & 58878.2058(1)  & 58902.3519(3)  &  59421.5030(10) & 59444.3367(2)  & 59600.7660(10)  \\  
58699.2056(1)   & 58722.3024(1)  & 58878.4677(2)  & 58902.6069(6)  &  59421.7650(20) & 59444.5996(1)  & 59601.0278(2)   \\  
58699.4674(2)   & 58722.5642(2)  & 58878.7308(1)  & 58902.8732(8)  &  59422.0278(1)  & 59444.8616(2)  & 59601.2894(2)   \\  
58699.7303(1)   & 58722.8273(1)  & 58878.9927(2)  & 58903.1397(6)  &  59422.2898(2)  & 59445.1246(1)  & 59601.5529(2)   \\  
58699.9923(2)   & 58723.0893(2)  & 58879.2555(1)  & 58903.4005(9)  &  59422.5527(1)  & 59445.3865(2)  & 59601.8143(2)   \\  
58700.2553(1)   & 58723.3520(12) & 58879.5176(2)  & 58903.6752(8)  &  59422.8147(2)  & 59445.6495(1)  & 59602.0777(2)   \\  
58700.5173(2)   & 58723.6155(2)  & 58879.7806(1)  & 58903.9267(4)  &  59423.0777(1)  & 59445.9114(2)  & 59602.3393(2)   \\  
58700.7803(1)   & 58725.1890(23) & 58880.0425(2)  & 58904.1896(1)  &  59423.3397(2)  & 59446.1744(1)  & 59602.6026(2)   \\  
58701.0422(2)   & 58725.4521(1)  & 58880.3056(1)  & 58904.4516(4)  &  59423.6027(1)  & 59446.4364(2)  & 59602.8642(2)   \\  
58701.3053(1)   & 58725.7154(2)  & 58880.5674(2)  & 58904.7147(1)  &  59423.8646(2)  & 59580.0305(1)  & 59603.1274(2)   \\  
58701.5570(11)  & 58725.9739(9)  & 58880.8305(1)  & 58904.9773(3)  &  59424.1275(1)  & 59580.2920(20) & 59603.3891(2)   \\  
58701.8292(3)   & 58726.2402(1)  & 58881.0923(2)  & 58905.2393(1)  &  59424.3894(2)  & 59580.5558(2)  & 59603.6525(2)   \\  
58702.0927(2)   & 58726.5019(1)  & 58881.3554(1)  & 58905.5015(4)  &  59424.6525(1)  & 59580.8172(2)  & 59603.9141(2)   \\  
58702.3552(1)   & 58726.7639(2)  & 58881.6173(2)  & 58906.0240(9)  &  59424.9144(2)  & 59581.0806(2)  & 59604.1773(2)   \\  
58702.6170(20)  & 58727.0268(1)  & 58881.8804(1)  & 58906.5460(22) &  59425.1773(1)  & 59581.3422(2)  & 59604.4390(20)  \\  
58702.8797(1)   & 58727.2888(2)  & 58882.1422(2)  & 58906.8143(7)  &  59425.4394(2)  & 59581.6057(2)  & 59604.7026(1)   \\  
58703.1419(2)   & 58727.5517(1)  & 58882.4053(1)  & 58907.0720(11) &  59425.7022(1)  & 59581.8670(20) & 59604.9638(2)   \\  
58703.4050(1)   & 58727.8136(2)  & 58882.6668(1)  & 58907.3392(6)  &  59425.9640(10) & 59582.1305(2)  & 59605.2275(2)   \\  
58703.6669(2)   & 58728.0766(1)  & 58882.9380(10) & 58907.5960(22) &  59426.2276(7)  & 59582.3920(20) & 59605.4888(2)   \\  
58703.9298(1)   & 58728.3386(2)  & 58883.1923(2)  & 58907.8639(6)  &  59426.4890(20) & 59582.6552(2)  & 59605.7526(2)   \\  
58704.1918(2)   & 58728.6015(1)  & 58883.4553(1)  & 58908.1210(19) &  59426.7521(1)  & 59582.9169(2)  & 59606.0138(2)   \\  
58704.4548(1)   & 58728.8633(2)  & 58883.7169(1)  & 58908.3891(6)  &  59427.0141(2)  & 59583.1802(2)  & 59606.2774(2)   \\  
58704.7168(2)   & 58729.1258(4)  & 58885.2924(1)  & 58908.6470(10) &  59427.2771(1)  & 59583.4419(2)  & 59606.5387(2)   \\  
58704.9797(1)   & 58729.3882(3)  & 58885.5549(1)  & 58909.1720(11) &  59427.5390(19) & 59583.7054(2)  &                 \\  
58705.2418(2)   & 58729.6514(1)  & 58885.8167(2)  & 58909.6960(22) &  59427.8019(1)  & 59583.9668(2)  &                 \\  
58705.5044(1)   & 58729.9134(2)  & 58886.0798(1)  & 58909.9638(6)  &  59428.0640(20) & 59584.2301(2)  &                 \\  
58705.7660(1)   & 58730.1763(1)  & 58886.3415(2)  & 58910.2220(10) &  59428.3270(10) & 59584.4918(2)  &                 \\  
58706.0289(2)   & 58730.4383(2)  & 58886.6049(1)  & 58910.4888(7)  &  59428.5889(2)  & 59584.7552(2)  &                 \\  

\end{longtable}
\tiny
\begin{longtable}{|l|l|l|l|l|l|l|}
\caption{VY UMi times of minima determined from TESS lightcurves. Errors are in parenthesis.} 
\label{tab:tmin_VY_UMi } \\
\hline 

\textbf{BJD} & 
\textbf{BJD} & 
\textbf{BJD} & 
\textbf{BJD} & 
\textbf{BJD} &
\textbf{BJD} &
\textbf{BJD}\\
\hline 

\endfirsthead

\multicolumn{7}{c}%
{{\bfseries \tablename\ \thetable{} -- continued from previous page}} \\
\hline 
\textbf{BJD} & 
\textbf{BJD} & 
\textbf{BJD} & 
\textbf{BJD} & 
\textbf{BJD} &
\textbf{BJD} &
\textbf{BJD}\\
\hline 
\endhead

\hline \multicolumn{7}{|r|}{{Continued on next page}} \\ \hline
\endfoot

\hline \hline
\endlastfoot

59390.7777(4) &  59406.5601(1)  & 59423.1556(1)  & 59439.1004(1)  & 59586.9968(2)  & 59602.7798(1)  & 59619.5379(2) \\
59390.9408(1) &  59406.7234(1)  & 59423.3189(1)  & 59439.2638(1)  & 59587.1606(1)  & 59602.9422(1)  & 59619.7013(1) \\
59391.1038(1) &  59406.8855(1)  & 59423.4809(1)  & 59439.4256(1)  & 59587.3229(2)  & 59603.1054(1)  & 59619.8635(2) \\
59391.2662(9) &  59407.0486(1)  & 59423.6442(1)  & 59439.5892(1)  & 59587.4858(1)  & 59603.2671(1)  & 59620.0265(1) \\
59391.4289(2) &  59407.2108(1)  & 59423.8064(1)  & 59439.7511(1)  & 59587.6475(2)  & 59603.4307(1)  & 59620.1888(2) \\
59391.5915(1) &  59407.3742(6)  & 59423.9697(1)  & 59439.9134(6)  & 59587.8113(1)  & 59603.5931(2)  & 59620.3517(1) \\
59391.7547(1) &  59407.5363(1)  & 59424.1318(1)  & 59440.0763(3)  & 59587.9738(2)  & 59603.7562(1)  & 59620.5133(2) \\
59391.9169(1) &  59407.6994(1)  & 59424.2951(1)  & 59440.2385(5)  & 59588.1367(1)  & 59603.9184(1)  & 59620.6776(1) \\
59392.0801(1) &  59407.8616(1)  & 59424.4572(1)  & 59440.4021(1)  & 59588.2990(20) & 59604.0816(1)  & 59620.8388(2) \\
59392.2424(1) &  59408.0249(1)  & 59424.6205(1)  & 59440.5653(1)  & 59588.4624(1)  & 59604.2432(2)  & 59621.0027(1) \\
59392.4057(1) &  59408.1872(1)  & 59424.7823(1)  & 59440.7273(1)  & 59588.6246(2)  & 59604.4070(10) & 59621.1641(2) \\
59392.5679(1) &  59408.3502(1)  & 59424.9458(1)  & 59440.8908(4)  & 59588.7877(1)  & 59604.5693(2)  & 59621.3282(1) \\
59392.7308(1) &  59408.5126(1)  & 59425.1076(1)  & 59441.0528(1)  & 59588.9499(2)  & 59604.7324(1)  & 59623.7672(3) \\
59392.8932(1) &  59408.6759(1)  & 59425.2713(1)  & 59441.2162(1)  & 59589.1131(1)  & 59604.8948(2)  & 59623.9314(1) \\
59393.0563(1) &  59408.8379(1)  & 59425.4330(1)  & 59441.3783(8)  & 59589.2746(2)  & 59605.0579(1)  & 59624.0937(2) \\
59393.2186(1) &  59409.0012(1)  & 59425.5967(1)  & 59441.5415(1)  & 59589.4384(1)  & 59605.2201(2)  & 59624.2567(1) \\
59393.3817(1) &  59409.1633(7)  & 59425.7585(1)  & 59441.7035(1)  & 59589.6001(2)  & 59605.3833(1)  & 59624.4181(2) \\
59393.5440(1) &  59409.3265(5)  & 59425.9221(1)  & 59441.8670(1)  & 59589.7640(10) & 59605.5454(1)  & 59624.5821(1) \\
59393.7071(1) &  59409.4887(1)  & 59426.0814(7)  & 59442.0290(1)  & 59589.9254(2)  & 59605.7087(1)  & 59624.7435(2) \\
59393.8695(1) &  59409.8141(8)  & 59426.2465(6)  & 59442.1924(1)  & 59590.0894(1)  & 59605.8704(1)  & 59624.9080(10)\\
59394.0326(1) &  59409.9773(1)  & 59426.4062(7)  & 59442.3543(9)  & 59590.2510(20) & 59606.0341(1)  & 59625.0699(2) \\
59394.1949(1) &  59410.1395(1)  & 59426.5724(2)  & 59442.5177(1)  & 59590.4148(1)  & 59606.1958(1)  & 59625.2333(1) \\
59394.3579(6) &  59410.3029(1)  & 59426.7349(1)  & 59442.6798(1)  & 59590.5763(2)  & 59606.3593(1)  & 59625.3954(2) \\
59394.5203(1) &  59410.4649(1)  & 59426.8984(1)  & 59442.8432(1)  & 59590.7402(1)  & 59606.5210(21) & 59625.5582(1) \\
59394.6834(1) &  59410.6282(1)  & 59427.0602(1)  & 59443.0053(8)  & 59590.9025(2)  & 59606.6845(1)  & 59625.7198(2) \\
59394.8458(1) &  59410.7903(1)  & 59427.2239(2)  & 59443.1686(1)  & 59591.0655(1)  & 59606.8457(6)  & 59625.8842(1) \\
59395.0089(1) &  59410.9536(1)  & 59427.3855(8)  & 59443.3307(1)  & 59591.2270(22) & 59608.7994(5)  & 59626.0452(2) \\
59395.1711(1) &  59411.1158(8)  & 59427.5491(1)  & 59443.4936(2)  & 59591.3909(1)  & 59608.9619(6)  & 59626.2091(1) \\
59395.3340(1) &  59411.2790(1)  & 59427.7110(1)  & 59443.6561(1)  & 59591.5524(2)  & 59609.1250(20) & 59626.3716(2) \\
59395.4964(1) &  59411.4411(1)  & 59427.8745(1)  & 59443.8194(1)  & 59591.7167(1)  & 59609.2881(1)  & 59626.5349(1) \\
59395.6595(1) &  59411.6045(1)  & 59428.0365(1)  & 59443.9814(1)  & 59591.8786(2)  & 59609.4504(2)  & 59626.6960(20)\\
59395.8218(1) &  59411.7662(1)  & 59428.1999(1)  & 59444.1450(20) & 59592.0418(1)  & 59609.6136(1)  & 59626.8599(1) \\
59395.9849(7) &  59411.9293(5)  & 59428.3618(1)  & 59444.3068(1)  & 59592.2041(2)  & 59609.7759(2)  & 59627.0214(2) \\
59396.1472(1) &  59412.2538(5)  & 59428.5254(4)  & 59444.4702(1)  & 59592.3671(1)  & 59609.9389(1)  & 59627.1854(1) \\
59396.3105(6) &  59412.4174(1)  & 59428.6871(1)  & 59444.6322(1)  & 59592.5294(2)  & 59610.1014(2)  & 59627.3469(2) \\
59396.4726(1) &  59412.5807(1)  & 59428.8502(2)  & 59444.7956(4)  & 59592.6923(1)  & 59610.2643(1)  & 59627.5107(1) \\
59396.6359(1) &  59412.7423(1)  & 59429.0128(1)  & 59444.9575(1)  & 59592.8531(7)  & 59610.4258(2)  & 59627.6733(2) \\
59396.7979(1) &  59412.9060(1)  & 59429.1762(1)  & 59445.1211(1)  & 59594.1564(2)  & 59610.5896(10) & 59627.8365(1) \\
59396.9612(1) &  59413.0683(1)  & 59429.3381(1)  & 59445.2829(1)  & 59594.3195(1)  & 59610.7512(2)  & 59627.9986(3) \\
59397.1235(1) &  59413.2315(1)  & 59429.5013(1)  & 59445.4465(1)  & 59594.4818(2)  & 59610.9151(1)  & 59628.1617(1) \\
59397.2868(8) &  59413.3933(1)  & 59429.6636(1)  & 59445.6085(1)  & 59594.6448(1)  & 59611.0776(2)  & 59628.3240(20)\\
59397.4484(8) &  59413.5568(1)  & 59429.8271(1)  & 59445.7719(1)  & 59594.8071(2)  & 59611.2406(1)  & 59628.4874(1) \\
59397.6121(5) &  59413.7187(1)  & 59429.9891(1)  & 59445.9338(1)  & 59594.9701(1)  & 59611.4029(2)  & 59628.6495(2) \\
59397.7769(8) &  59413.8823(1)  & 59430.1524(1)  & 59446.0972(1)  & 59595.1324(1)  & 59611.5659(1)  & 59628.8123(1) \\
59397.9374(1) &  59414.0444(1)  & 59430.3143(1)  & 59446.2594(1)  & 59595.2957(1)  & 59611.7274(2)  & 59628.9748(3) \\
59398.0996(1) &  59414.2076(1)  & 59430.4778(6)  & 59446.4218(3)  & 59595.4574(1)  & 59611.8914(1)  & 59629.1382(1) \\
59398.2627(1) &  59414.3698(1)  & 59430.6397(1)  & 59579.8384(5)  & 59595.6210(1)  & 59612.0537(2)  & 59629.3015(7) \\
59398.4249(1) &  59414.5330(1)  & 59430.8032(1)  & 59580.0012(5)  & 59595.7833(2)  & 59612.2168(1)  & 59629.6234(8) \\
59398.5883(1) &  59414.6951(1)  & 59430.9652(1)  & 59580.1634(2)  & 59595.9464(1)  & 59612.3791(2)  & 59629.7886(1) \\
59398.7504(1) &  59414.8587(2)  & 59431.1286(1)  & 59580.3274(1)  & 59596.1089(2)  & 59612.5422(1)  & 59630.1141(1) \\
59398.9137(1) &  59415.0204(1)  & 59431.2905(1)  & 59580.4896(2)  & 59596.2717(1)  & 59612.7036(2)  & 59630.2754(2) \\
59399.0757(1) &  59415.1837(1)  & 59431.4540(1)  & 59580.6528(1)  & 59596.4342(1)  & 59612.8676(1)  & 59630.4395(1) \\
59399.2389(1) &  59415.5093(1)  & 59431.6159(1)  & 59580.8150(20) & 59596.5971(1)  & 59613.0299(2)  & 59630.6018(2) \\
59399.4011(1) &  59415.6712(1)  & 59431.7795(1)  & 59580.9779(1)  & 59596.7596(1)  & 59613.1929(1)  & 59630.7651(1) \\
59399.5645(1) &  59415.8346(1)  & 59431.9413(1)  & 59581.1396(2)  & 59596.9225(1)  & 59613.3554(2)  & 59630.9274(2) \\
59399.7266(1) &  59415.9966(1)  & 59432.1048(1)  & 59581.3035(1)  & 59597.0850(21) & 59613.5184(1)  & 59631.0902(1) \\
59399.8894(1) &  59416.1601(1)  & 59432.2669(1)  & 59581.4649(2)  & 59597.2479(1)  & 59613.6799(2)  & 59631.2528(3) \\
59400.0519(9) &  59416.3219(1)  & 59432.4311(4)  & 59581.6291(1)  & 59597.4104(1)  & 59613.8436(8)  & 59631.4158(1) \\
59400.2153(1) &  59416.4851(2)  & 59433.8933(5)  & 59581.7912(2)  & 59597.5735(1)  & 59614.0052(3)  & 59631.5770(29)\\
59400.3774(1) &  59416.6475(1)  & 59434.0574(1)  & 59581.9541(1)  & 59597.7352(1)  & 59614.1695(1)  & 59631.7413(1) \\
59400.5406(1) &  59416.8109(1)  & 59434.2193(1)  & 59582.1166(2)  & 59597.8988(1)  & 59614.3307(3)  & 59631.9035(2) \\
59400.7028(1) &  59416.9728(8)  & 59434.3823(2)  & 59582.2799(1)  & 59598.0612(2)  & 59614.4938(2)  & 59632.0666(1) \\
59401.0281(1) &  59417.1365(2)  & 59434.5448(1)  & 59582.4412(2)  & 59598.2242(1)  & 59614.6582(4)  & 59632.2277(3) \\
59401.1915(1) &  59417.2979(1)  & 59434.7082(1)  & 59582.6052(1)  & 59598.3865(1)  & 59615.1456(1)  & 59632.3921(1) \\
59401.3535(1) &  59417.4616(1)  & 59434.8701(1)  & 59582.7674(2)  & 59598.5496(9)  & 59615.3078(2)  & 59632.5532(3) \\
59401.5168(1) &  59417.6237(1)  & 59435.0337(1)  & 59582.9306(1)  & 59598.7114(1)  & 59615.4707(1)  & 59632.7176(1) \\
59401.6789(1) &  59417.7871(1)  & 59435.1954(1)  & 59583.0928(2)  & 59598.8751(1)  & 59615.6333(2)  & 59632.8798(2) \\
59401.8423(1) &  59417.9489(1)  & 59435.3589(1)  & 59583.2558(1)  & 59599.0375(2)  & 59615.7960(10) & 59633.0428(1) \\
59402.0044(1) &  59418.1126(2)  & 59435.5209(1)  & 59583.4182(2)  & 59599.2005(1)  & 59615.9575(2)  & 59633.2052(3) \\
59402.1675(1) &  59418.2743(1)  & 59435.6843(1)  & 59583.5813(1)  & 59599.3628(1)  & 59616.1219(1)  & 59633.3679(1) \\
59402.3297(1) &  59418.4379(1)  & 59435.8461(1)  & 59583.7427(2)  & 59599.5258(1)  & 59616.2839(2)  & 59633.5306(3) \\
59402.4929(1) &  59418.5990(20) & 59436.0097(1)  & 59583.9066(1)  & 59599.6876(2)  & 59616.4469(1)  & 59633.6938(1) \\
59402.6551(1) &  59420.2257(4)  & 59436.1718(1)  & 59584.0691(2)  & 59599.8513(1)  & 59616.6093(2)  & 59633.8548(2) \\
59402.8185(1) &  59420.3904(1)  & 59436.3351(1)  & 59584.2322(1)  & 59600.0136(1)  & 59616.7726(1)  & 59634.0191(1) \\
59402.9807(1) &  59420.5522(1)  & 59436.4972(1)  & 59584.3944(2)  & 59600.1756(3)  & 59616.9348(2)  & 59634.1814(3) \\
59403.1442(2) &  59420.7157(4)  & 59436.6606(1)  & 59584.5575(1)  & 59600.3387(4)  & 59617.0980(10) & 59634.3442(1) \\
59403.3060(1) &  59420.8776(1)  & 59436.8224(1)  & 59584.7199(2)  & 59600.5006(5)  & 59617.2602(2)  & 59634.5068(3) \\
59403.4694(1) &  59421.0412(1)  & 59436.9859(1)  & 59584.8830(10) & 59600.6648(5)  & 59617.4234(1)  & 59634.6695(1) \\
59403.6312(1) &  59421.2028(1)  & 59437.1478(1)  & 59585.0452(2)  & 59600.8274(1)  & 59617.5856(2)  & 59634.8311(2) \\
59403.7945(1) &  59421.3664(1)  & 59437.3113(1)  & 59585.2083(1)  & 59600.9891(2)  & 59617.7485(9)  & 59634.9950(10)\\
59403.9568(1) &  59421.5285(1)  & 59437.4734(1)  & 59585.3706(2)  & 59601.1529(1)  & 59617.9110(20) & 59635.1576(2) \\
59404.1201(1) &  59421.6919(1)  & 59437.6367(1)  & 59585.5337(1)  & 59601.3146(2)  & 59618.0741(1)  & 59635.3204(1) \\
59404.2819(8) &  59421.8538(1)  & 59437.7987(1)  & 59585.6960(21) & 59601.4783(1)  & 59618.2355(2)  & 59635.4831(3) \\
59405.4201(5) &  59422.0172(1)  & 59437.9621(1)  & 59585.8591(1)  & 59601.6406(2)  & 59618.3995(1)  & 59635.6457(1) \\
59405.5834(1) &  59422.1790(1)  & 59438.1242(1)  & 59586.0213(2)  & 59601.8038(1)  & 59618.5617(2)  & 59635.8090(10)\\
59405.7470(1) &  59422.3428(1)  & 59438.2875(1)  & 59586.1834(5)  & 59601.9651(2)  & 59618.7251(1)  &               \\
59405.9091(1) &  59422.5048(1)  & 59438.4496(1)  & 59586.3450(28) & 59602.1290(1)  & 59618.8873(2)  &               \\
59406.0724(6) &  59422.6680(1)  & 59438.6129(1)  & 59586.5115(5)  & 59602.2913(1)  & 59619.0501(1)  &               \\
59406.2347(1) &  59422.8298(1)  & 59438.7748(1)  & 59586.6721(2)  & 59602.4545(9)  & 59619.2117(2)  &               \\
59406.3979(1) &  59422.9935(1)  & 59438.9384(1)  & 59586.8353(1)  & 59602.6169(2)  & 59619.3758(1)  &               \\
\end{longtable}

\tiny
\begin{longtable}{|l|l|l|l|l|l|l|}
\caption{GSC 04364-0064 times of minima determined from TESS lightcurves. Errors are in parenthesis.} 
\label{tab:tmin_GSC} \\
\hline 

\textbf{BJD} & 
\textbf{BJD} & 
\textbf{BJD} & 
\textbf{BJD} & 
\textbf{BJD} &
\textbf{BJD} &
\textbf{BJD}\\
\hline 

\endfirsthead

\multicolumn{7}{c}%
{{\bfseries \tablename\ \thetable{} -- continued from previous page}} \\
\hline 
\textbf{BJD} & 
\textbf{BJD} & 
\textbf{BJD} & 
\textbf{BJD} & 
\textbf{BJD} &
\textbf{BJD} &
\textbf{BJD}\\
\hline 
\endhead

\hline \multicolumn{7}{|r|}{{Continued on next page}} \\ \hline
\endfoot

\hline \hline
\endlastfoot

58842.925(2) & 58848.963(1) & 58851.983(4) & 58858.454(2) &  58864.494(2) & 59014.199(3) & 59019.375(4) \\
58843.356(4) & 58849.391(4) & 58851.985(3) & 58858.886(3) &  58864.926(3) & 59014.632(1) & 59019.809(2) \\
58843.786(1) & 58849.394(4) & 58852.414(1) & 58859.749(4) &  58865.789(3) & 59014.634(1) & 59020.241(3) \\
58845.512(2) & 58849.826(2) & 58852.415(2) & 58860.609(4) &  58867.085(2) & 59015.498(2) & 59020.672(1) \\
58845.513(1) & 58849.827(1) & 58852.844(4) & 58861.473(4) &  59011.179(2) & 59015.923(4) & 59020.674(2) \\
58845.937(4) & 58850.259(4) & 58853.709(4) & 58862.338(4) &  59012.045(2) & 59016.789(4) & 59021.534(2) \\
58845.945(4) & 58850.687(2) & 58854.140(8) & 58862.767(2) &  59012.476(4) & 59017.648(4) & 59026.712(1) \\
58846.374(2) & 58850.688(2) & 58856.727(2) & 58862.768(2) &  59012.906(2) & 59018.085(2) & 59029.729(4) \\
58846.376(2) & 58851.119(4) & 58857.158(4) & 58863.199(4) &  59012.907(2) & 59018.086(2) & 59031.888(2) \\
58846.804(4) & 58851.552(1) & 58857.593(2) & 58863.634(2) &  59013.335(4) & 59018.514(3) & 59033.614(2) \\
58848.101(2) & 58851.553(2) & 58858.022(4) & 58864.065(4) &  59013.773(2) & 59018.946(2) & 59034.481(2) \\

\end{longtable}

\tiny
\begin{longtable}{|l|l|l|l|}
\caption{Observed times of minima of selected EB systems. Errors are in parenthesis.} 
\label{tab:tmin_our} \\
\hline 

\textbf{Name} & 
\textbf{BJD} &
\textbf{Name} & 
\textbf{BJD}\\
\hline 

\endfirsthead

\multicolumn{4}{c}%
{{\bfseries \tablename\ \thetable{} -- continued from previous page}} \\
\hline 
\textbf{Name} & 
\textbf{BJD} &
\textbf{Name} & 
\textbf{BJD}\\
\hline 
\endhead

\hline \multicolumn{4}{|r|}{{Continued on next page}} \\ \hline
\endfoot

\hline \hline
\endlastfoot

RU UMi & 59277.4101(2) &  VY UMi & 59516.3846(2)         \\
RU UMi & 59467.4327(1) &  VY UMi & 59517.3604(1)         \\
VY UMi & 59298.3633(2) &  VY UMi & 59517.5219(1)         \\
VY UMi & 59298.5245(2) &  GSC 04364-0064  & 59343.3781(2)\\
VY UMi & 59516.2200(3) &  GSC 04364-0064  & 59374.4401(1)\\

\end{longtable}

\end{appendices}
\end{document}